\documentclass[twocolumn,amssymb,dvipdfmx]{revtex4-1}%
 
\usepackage{amsmath}
\usepackage{ascmac} 
\usepackage{ascmac}
\usepackage{color}
\usepackage{amsmath,amssymb} 
\usepackage{braket}
\usepackage[pdftex]{graphicx}
 
\begin{document}

\title{Enhancing quantum annealing performance by a degenerate two-level system}

\author{Shohei Watabe$^{1,2}$, Yuya Seki$^2$, and Shiro Kawabata$^2$}  
\affiliation{$^1$ Department of Physics, Faculty of Science Division I, Tokyo University of Science, Shinjuku, Tokyo 162-8601, Japan.}
\affiliation{$^2$ Nanoelectronics Research Institute, National Institute of Advanced Industrial Science and Technology (AIST),
1-1-1 Umezono, Tsukuba, Ibaraki 305-8568, Japan.}


\begin{abstract} 
Quantum annealing is an innovative idea and method for avoiding the increase of the calculation cost of the combinatorial optimization problem. Since the combinatorial optimization problems are ubiquitous, quantum annealing machine with high efficiency and scalability will give an immeasurable impact on many fields. However, the conventional quantum annealing machine may not have a high success probability for finding the solution because the energy gap closes exponentially as a function of the system size. To propose an idea for finding high success probability is one of the most important issues. 
Here we show that a degenerate two-level system provides the higher success probability than the conventional spin-1/2 model in a weak longitudinal magnetic field region. 
The physics behind this is that the quantum annealing in this model can be reduced into that in the spin-1/2 model, 
where the effective longitudinal magnetic field may open the energy gap, which suppresses the Landau--Zener tunneling providing leakage of the ground state. 
We also present the success probability of the $\Lambda$-type system, which may show the higher success probability than the conventional spin-1/2 model. 
\end{abstract}

\pacs{}

\maketitle


\section{Introduction}

Quantum annealing is an interesting approach for finding the optimal solution of combinatorial optimization problems by using the quantum effect~\cite{Kadowaki1998,Albash2018,Farhi2000,Farhi2001}.
The combinatorial optimization problems are ubiquitous in the real social world, therefore the spread of quantum annealing machine with high efficiency and high scalability will give impacts and benefits on many fields, such as an industry including drug design~\cite{Sakaguchi2016}, financial portfolio problem~\cite{Rosenberg2016}, and traffic flow optimization~\cite{Neukart2017}. 
After the commercialization of superconducting quantum annealing machine by D-Wave Systems inc.~\cite{D-waveURL}, hardware has been investigated and developed~\cite{Barends2016,Rosenberg2017,Novikov2018,Maezawa2019,Mukai2019}. 
    
However, there are bottlenecks for implementing scalable quantum annealing machine; for the conventional and scalable quantum annealing machine may not have a high success probability for finding the solution of a combinatorial optimization problem because of the emergence of the first order phase transition, where the energy gap between ground state and the first exited state closes exponentially as a function of the system size~\cite{Albash2018}. 
In this case, it necessitates an exponentially long annealing time for finding the solution of the problem~\cite{Znidaric2006,Jorg2010A,Jorg2010B}. 
In the case of the second oder phase transition, on the other hand, an annealing time for finding the solution may scales polynomially as a function of the system size~\cite{Seki2012}.

To propose an idea for finding high success probability is one of the most important and challenging issue in the field of quantum annealing. 
One of the approaches for obtaining the high success probability is to engineer the scheduling function for the driving Hamiltonian and the problem Hamiltonian, such as a monotonically increasing scheduling function satisfying the local adiabatic condition~\cite{Roland2002}, the reverse quantum annealing~\cite{Ortiz2011} implemented in D-wave 2000Q~\cite{reverse}, inhomogeneous sweeping out of local transverse magnetic fields~\cite{Susa2018,Susa2018PRA}, and a diabatic pulse application~\cite{Karanikolas2017}. 
Another is to add an artificial additional Hamiltonian for suppressing the emergence of the excitations with avoiding the slowing down of annealing time, 
which is called shortcuts to adiabaticity by the counterdiabatic driving~\cite{Campo2012,Campo2013,Sels2017,Hartmann2019}, and to add an additional Hamiltonian for avoiding the first order phase transition~\cite{Seki2012,Seoane2015,Seki2015}.  
In this paper, we study the possibility of other approach: to employ a variant spin, such as a qudit, in the quantum annealing architecture.

Recently, two of the authors have studied the quantum phase transition in a degenerate two-level spin system, called the quantum Wajnflasz--Pick model, where an internal spin state is coupled to all the same energy internal states with a single coupling strength, and to all the different energy internal states with the other single coupling strength~\cite{Seki2019}. 
In the earlier study, this model is found to show a several kinds of phase transition while annealing; single or double first-order phase transitions as well as a single second-order phase transition, depending on an internal state coupling parameter~\cite{Seki2019}, which suggests that the quantum annealing of this model may be controlled by an internal state tuning parameter. 
However, the study is based on the static statistical approach using the mean-field theory, because only the order of the phase transition has been interested in. 
Therefore, the enhancement of the success probability for quantum annealing based on degenerate two-level systems is not clear yet. 
Furthermore, they employed a fully-connected uniform interacting system, and it is unclear whether their idea works that a double (or even-number of) first-order phase transition while annealing would bring the system back into the ground state at the end of the annealing, where the even number of the Landau--Zener tunneling may happen with respect to the ground state. 

In the present paper, we clarify the success probability of the quantum annealing in the quantum Wajnflasz--Pick model, focusing on (i) the Schr\"odinger dynamics, (ii) eigenenergies, and (iii) {\it non-uniform} effects of the spin-interaction as well as the longitudinal magnetic field. 
We find that the quantum Wajnflasz--Pick model is more efficient than the conventional spin-1/2 model in the weak longitudinal magnetic field region as well as in the strong coupling region between degenerate states. 
We also find that the quantum Wajnflasz--Pick model is reducible into a spin-1/2 model, where effect of the transverse magnetic field in the original Hamiltonian emerges in the reduced Hamiltonian not only as the effective transverse magnetic field but also as the effective longitudinal magnetic field. 
As a result, this model may provide the higher success probability in the case where the effective longitudinal magnetic field opens the energy gap between the ground state and the first excited state. 
We also evaluate the success probability in another variant spin, a $\Lambda$-type system~\cite{Cirac1997,Duan2001,Zhou2002,Sun2003,Yang2003,Yang2004,Zhou2004,You2011,Falci2013,Inomata2014}, which has three internal levels. 
This model also shows the higher success probability than the conventional spin-1/2 model in the weak magnetic field region. 

A multilevel system is ubiquitous, which can be seen, for example, in degenerate two-level systems in atoms~\cite{Margalit2013,Zhang2019}, 
$\Lambda$-type atoms~\cite{Cirac1997,Duan2001,Sun2003}, 
$\Lambda$-, $V$-, $\Theta$- and $\Delta$-type systems in the superconducting circuits~\cite{Zhou2002,Yang2003,Yang2004,Zhou2004,You2011,Falci2013,Inomata2014,Liu2005} as well as $\Lambda$-type systems in the nitrogen-vacancy centre in diamond~\cite{Zhou2017}. 
We hope that insights of our results in the degenerate two-level system and knowledge of their reduced Hamiltonian inspire and promote further study as well as future engineering of quantum annealing.

\section{Quantum Wajnflasz--Pick model}

A conventional quantum annealing consists of the spin-$1/2$ model, where the time dependent Hamiltonian is given by~\cite{Kadowaki1998} 
\begin{align}
	\hat H (s) = s \hat H_z + \left ( 1 - s \right ) \hat H_x, 
	\label{eq1}
\end{align}
where $\hat H_{z,x}$ are a problem and driver Hamiltonian, respectively, and $s \equiv t/T$ is the time $t \in [0,T]$ scaled by the annealing time $T$. 
The problem Hamiltonian  $\hat H_{z}$ with the number of spins $N$, which encodes the desired optimal solution, has a non-trivial ground state. 
In contrast, the driver Hamiltonian $\hat H_{x}$ has a trivial ground state, where the driver Hamiltonian $\hat H_{x}$ must not be commutable with the problem Hamiltonian $\hat H_{z}$. 
A problem Hamiltonian and driver Hamiltonian are typically given by 
\begin{align}
	\hat H_z \equiv & - \sum\limits_{i \neq j }^N J_{ij} \hat \sigma_i^z \hat \sigma_j^z  - \sum\limits_i^N h_i^z  \hat \sigma_i^z, 
	\label{eq2_0}	
\\
	\hat H_x \equiv & - \sum\limits_i^N h_i^x \hat \sigma_i^x , 
	\label{eq3_0}	
\end{align}
where $\hat \sigma^{x,z}$ are the Pauli matrices, $J_{ij}$ is the coupling strength between spins, $h_i^z$ is the local longitudinal magnetic field, and $h_i^x$ is the local transverse magnetic field. 
The time-dependent total Hamiltonian $\hat H(s)$ gradually changes from the driver Hamiltonian $\hat H_x$ to the problem Hamiltonian $\hat H_z$. 
If the Hamiltonian changes sufficiently slowly, the quantum adiabatic theorem guarantees that the initial quantum ground state follows the instantaneous ground state of the total Hamiltonian~\cite{Messiah1976}. 
We can thus finally obtain a non-trivial ground state of the problem Hamiltonian starting from the trivial ground state of the deriver Hamiltonian making use of the Schr\"{o}dinger dynamics.

\begin{figure}
\begin{center}
\includegraphics[width=8cm]{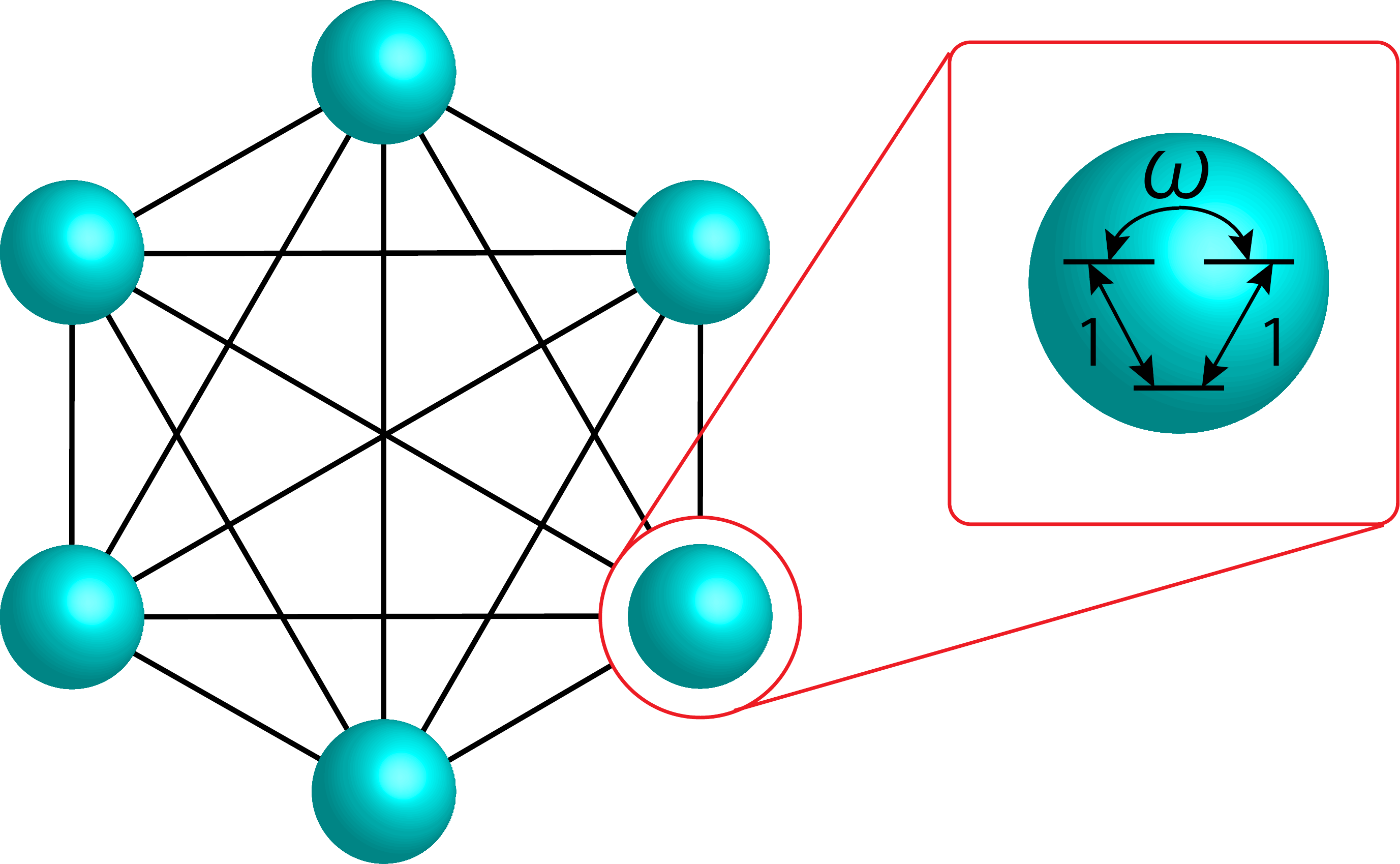}
\end{center}
\caption{Schematic setup of an interacting degenerate two-level system, called the quantum Wajnflasz--Pick model. }
\label{fig1.fig}
\end{figure}

The quantum Wajnflasz--Pick model is a quantum version of the Wajnflasz--Pick model~\cite{Wajnflasz1971}, which can describe one of the interacting degenerate two-level systems. In the language of the quantum annealing, 
the problem Hamiltonian and the driver Hamiltonian are respectively given by~\cite{Seki2019} 
\begin{align}
	\hat H_z \equiv & - \sum\limits_{i \neq j }^N J_{ij} \hat \tau_i^z \hat \tau_j^z  - \sum\limits_i^N h_i^z  \hat \tau_i^z, 
	\label{eq2}	
\\ 
	\hat H_x \equiv & - \sum\limits_i^N h_i^x \hat \tau_i^x. 
	\label{eq3}	
\end{align} 
(Schematic picture of this model is shown in Fig.~\ref{fig1.fig}.)
The Hamiltonian of this model can be simply obtained by replacing the Pauli matrices $\hat \sigma^{x,z}$ in Eqs.~\eqref{eq2_0} and~\eqref{eq3_0} with the spin matrices of the quantum Wajnflasz-Pick model $\hat \tau^{x,z}$. 
The spin operator $\hat \tau^z$ is given by~\cite{Seki2019} 
\begin{align}
	\hat \tau^z \equiv {\rm diag}  ( \underbrace{+1, \cdots, +1}_{g_{\rm u}}, \underbrace{-1, \cdots, -1}_{g_{\rm l}}), 
	\label{eqtauz}
\end{align}
where $g_{\rm u(l)}$ is the number of the degeneracy of the upper (lower) states.  
The spin-operator $\hat \tau^x$ in the driver Hamiltonian is given by 
\begin{align}
	\hat \tau^x \equiv 
	\frac{1}{c} 
	\begin{pmatrix} 
	{\bf A} (g_{\rm u}) & {\bf 1} (g_{\rm u}, g_{\rm l}) 
	\\
	{\bf 1} (g_{\rm l}, g_{\rm u}) & {\bf A} (g_{\rm l})
	\end{pmatrix}, 
	\label{eqtaux}	
\end{align}
where ${\bf A} (l)$ is a $(l\times l)$ matrix with the off-diagonal term $\omega$, given by 
\begin{align}
	{\bf A} (l) \equiv 
	\begin{pmatrix} 
	0 & \omega & \cdots & \omega
	\\
	\omega^* & 0 & \ddots & \vdots
	\\ 
	\vdots & \ddots & \ddots & \omega
	\\
	\omega^* & \cdots & \omega^* & 0
	\end{pmatrix}. 
\end{align}
Here, $\omega$ is a parameter of the internal transition between the degenerated upper/lower states. 
The matrix ${\bf 1} (m,n)$ is the $(m \times n)$ matrix, where all the elements is unity, which gives the transition between the upper and lower states.  
The constant $c$ is the normalization factor, where the maximum eigenvalue is normalized to be $+1$, so as to be equal to the maximum eigenvalue of $\hat \tau^z$. 

In the following, for the consistency to the earlier work~\cite{Seki2019}, we consider a uniform transverse field $h_i^x \equiv 1$, and also take the parameter of the internal transition to be real $\omega = \omega^*$ with $\omega > -1$. 
In the case where $(g_{\rm u}, g_{\rm l}) = (2,1)$, we have 
\begin{align}
	\hat \tau^z \equiv 
	\begin{pmatrix} 
	1 & 0 & 0 
	\\ 
	0 & 1 & 0 
	\\ 
	0 & 0 & - 1
	\end{pmatrix}, 
	\quad 
	\hat \tau^x \equiv 
	\frac{1}{c}
	\begin{pmatrix} 
	0 & \omega & 1
	\\ 
	\omega & 0 & 1
	\\ 
	1 & 1 & 0
	\end{pmatrix}, 	
	\label{matrixtauzx}
\end{align}
with $c = (\omega + \sqrt{8 + \omega^2})/2$, which is a kind of the $\Delta$-type system~\cite{You2011}. 

In this paper, we employ the common sets of parameters in both quantum Wajnflasz--Pick model and the conventional spin-$1/2$ model, including the coupling strength $J_{ij}$, the magnetic fields $h_i^{z,x}$, and the annealing time $T$. By using these parameters, we can obtain the same spin configuration ($+1$ or $-1$) in the ground state of the problem Hamiltonian. We thus compare efficiency of these models from the success probability.

\section{Schr\"odinger dynamics} 

In order to numerically calculate the success probability of the quantum annealing, 
we employ the Crank--Nicholson method~\cite{Press2002} for solving the Schr\"odinger equation 
\begin{align}
	i \frac{d}{dt} | \Psi (t) \rangle = \hat H (t) | \Psi (t) \rangle. 
\end{align}
In this method, the time-evolution of the wavefunction is calculated by using the Cayley's form~\cite{Press2002} 
\begin{align}
	 | \Psi (t + \Delta t) \rangle = 
	\frac{1 - i \hat H  \Delta t/2}{1 + i \hat H  \Delta t/2}  | \Psi (t) \rangle. 
\end{align} 
Although the inverse matrix is needed, this method conserves the norm of the wave-function and is second-order accurate in time~\cite{Press2002}. 

\begin{figure}
\begin{center}
\includegraphics[width=8cm]{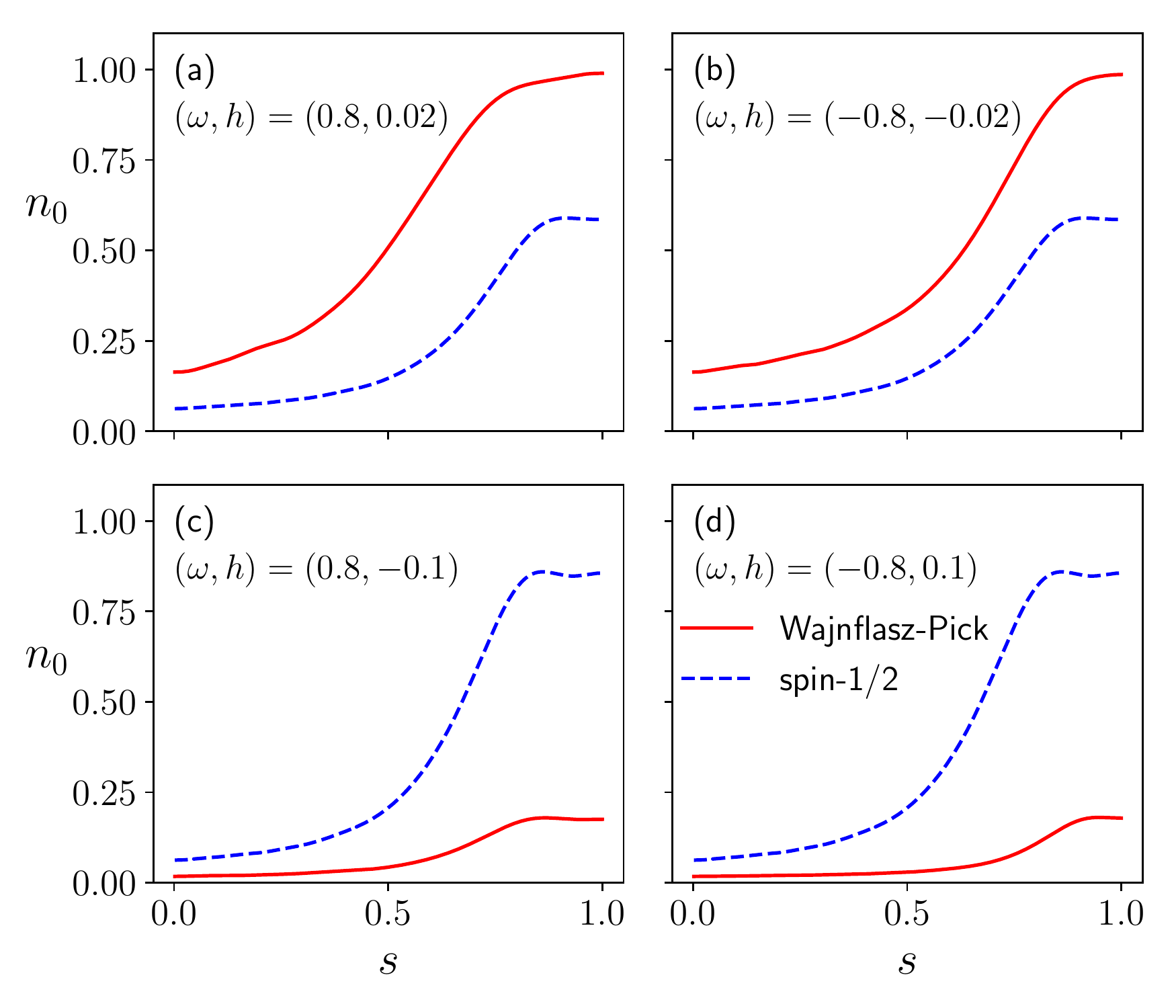}
\end{center}
\caption{Population $n_0$ of the ground state of the problem Hamiltonian $\hat H_z$ in the quantum Wajnflasz--Pick model and that of the conventional spin-1/2 model, where $n_0 \equiv | \langle \Psi (t) | \Psi_0 (T) \rangle |^2$. The scaled time $s$ is given by $s\equiv t/T$. 
We employed the number of spins $N=4$ both in the quantum Wajnflasz--Pick model and in the spin-1/2 model. We used the parameters $(g_{\rm u}, g_{\rm l}) = (2,1)$, $J_{ij} = 1/N$, $h_i^x = 1$ and $T=10$. 
}
\label{fig2.fig}
\end{figure}

We first consider the fully connected model, where the spin-spin coupling is ferromagnetic and the longitudinal magnetic field is uniform $h_i^z \equiv h$, which is consistent with the earlier work~\cite{Seki2019}. 
For example, in the case where $(\omega, h) = (0.8,0.02)$ and $(-0.8,-0.02)$ for $(g_{\rm u}, g_{\rm l}) = (2,1)$, 
the time-dependence of the ground state population of the problem Hamiltonian, given by $n_0 \equiv | \langle \Psi (t) | \Psi_0 (T) \rangle |^2$, clearly shows that this quantity in the quantum Wajnflasz--Pick model is greater than that in the conventional spin-1/2 model (Panels (a) and (b) in Fig.~\ref{fig2.fig}). 
Here, $|\Psi_0 (T) \rangle$ is the ground state of the problem Hamiltonian, and $| \Psi (t) \rangle$ is the wave function obtained from the time-dependent Schr\"odinger equation. 
In the case where $(\omega, h) = (0.8,-0.1)$ and $(-0.8,0.1)$ for $(g_{\rm u}, g_{\rm l}) = (2,1)$, on the other hand, the ground state population of the problem Hamiltonian in the quantum Wajnflasz--Pick model is less than that in the spin-1/2 model (Panels (c) and (d) in Fig.~\ref{fig2.fig}).  

\begin{figure}
\begin{center}
\includegraphics[width=8cm]{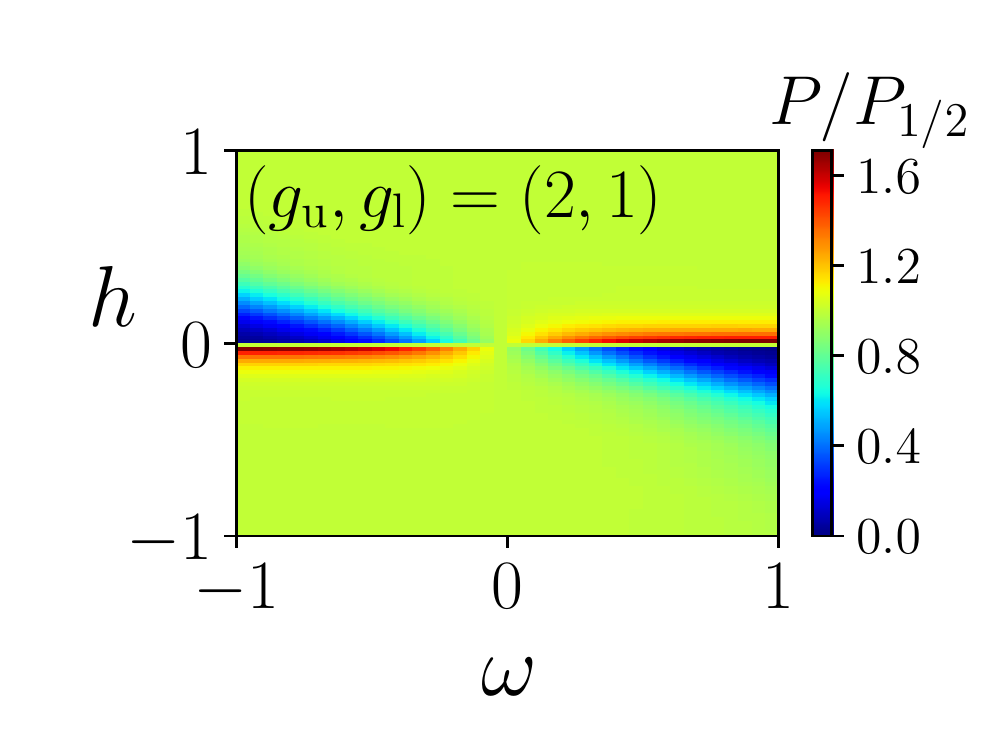}
\end{center}
\caption{Success probability of a quantum Wajnflasz--Pick model $P$ scaled by that of the conventional spin-1/2 model $P_{1/2}$ as a function of longitudinal magnetic field $h$ and the coupling strength $\omega$ between degenerate internal states. The parameters are the same as those used in Fig.~\ref{fig2.fig}. }
\label{fig3.fig}
\end{figure}

Compare the success probability of the quantum Wajnflasz--Pick model, $P \equiv | \langle \Psi (T) | \Psi_{0} (T) \rangle |^2$, with that of the conventional spin-1/2 model denoted as $P_{1/2}$, 
where $|\Psi (T) \rangle$ is the final state obtained from the time-dependent Schr\"odinger equation. 
In almost all regions in the $\omega$-$h$ plane, efficiencies of both models are almost the same, where the ratio of the success probability of the quantum Wajnflasz--Pick model and that of the conventional spin-1/2 model is almost unity (Fig.~\ref{fig3.fig}). 
On the other hand, in the regime of the weak longitudinal magnetic field $h$, we can find higher or lower efficiency regions in the quantum Wajnflasz--Pick model, compared with the spin-1/2 model. 
In the spin glass model, a non-trivial state may emerge in the weak longitudinal magnetic field limit~\cite{Nishimori2011}. 
In a $p$-spin model where $p=3,5,7, \cdots$, the energy gap is known to close exponentially and the first-order phase transition emerges in the absence of the longitudinal magnetic field~\cite{Jorg2010A}. 
In this sense, it is of interest that the quantum Wajnflasz--Pick model may provide the high efficiency in the weak longitudinal magnetic field region. 

\begin{figure}
\begin{center}
\includegraphics[width=7cm]{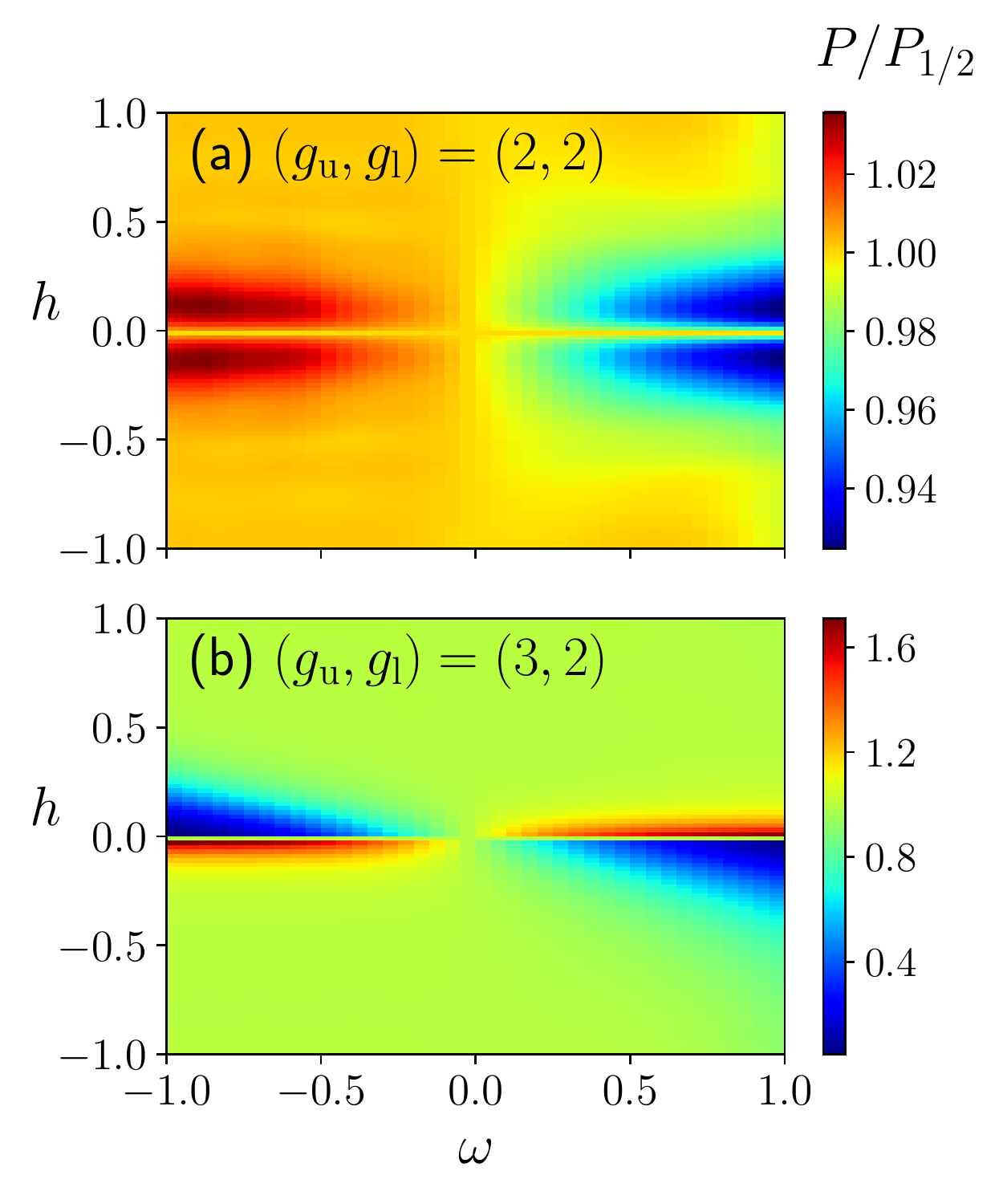}
\end{center}
\caption{Success probability of a quantum Wajnflasz--Pick model $P$ scaled by that of spin-1/2 model $P_{1/2}$. 
We consider the following degeneracy case: $(g_{\rm u}, g_{\rm l}) = (2,2)$ in Panels (a) and $(g_{\rm u}, g_{\rm l}) = (3,2)$ in Panels (b). 
We used $N=4$, $J_{ij} = 1/N$, $h_i^x = 1$ and $T=10$. 
} 
\label{fig4.fig}
\end{figure}

In the case where $(g_{\rm u}, g_{\rm l}) = (2,2)$, where the numbers of upper and lower states are equal, 
the success probability of the quantum Wajnflasz--Pick model is almost equal to that of the conventional spin-1/2 model (Panel (a) in Fig.~\ref{fig4.fig}). 
In the case where $(g_{\rm u}, g_{\rm l}) = (3,2)$, 
the success probability of the quantum Wajnflasz--Pick model is almost equal to that of the case where $(g_{\rm u}, g_{\rm l}) = (2,1)$, 
where the differences between the number of the upper states and that of lower states are the same in both cases (Fig.~\ref{fig3.fig} and Panel (b) in Fig.~\ref{fig4.fig}). 

\section{Eigenvalues}

\begin{figure}
\begin{center}
\includegraphics[width=8cm]{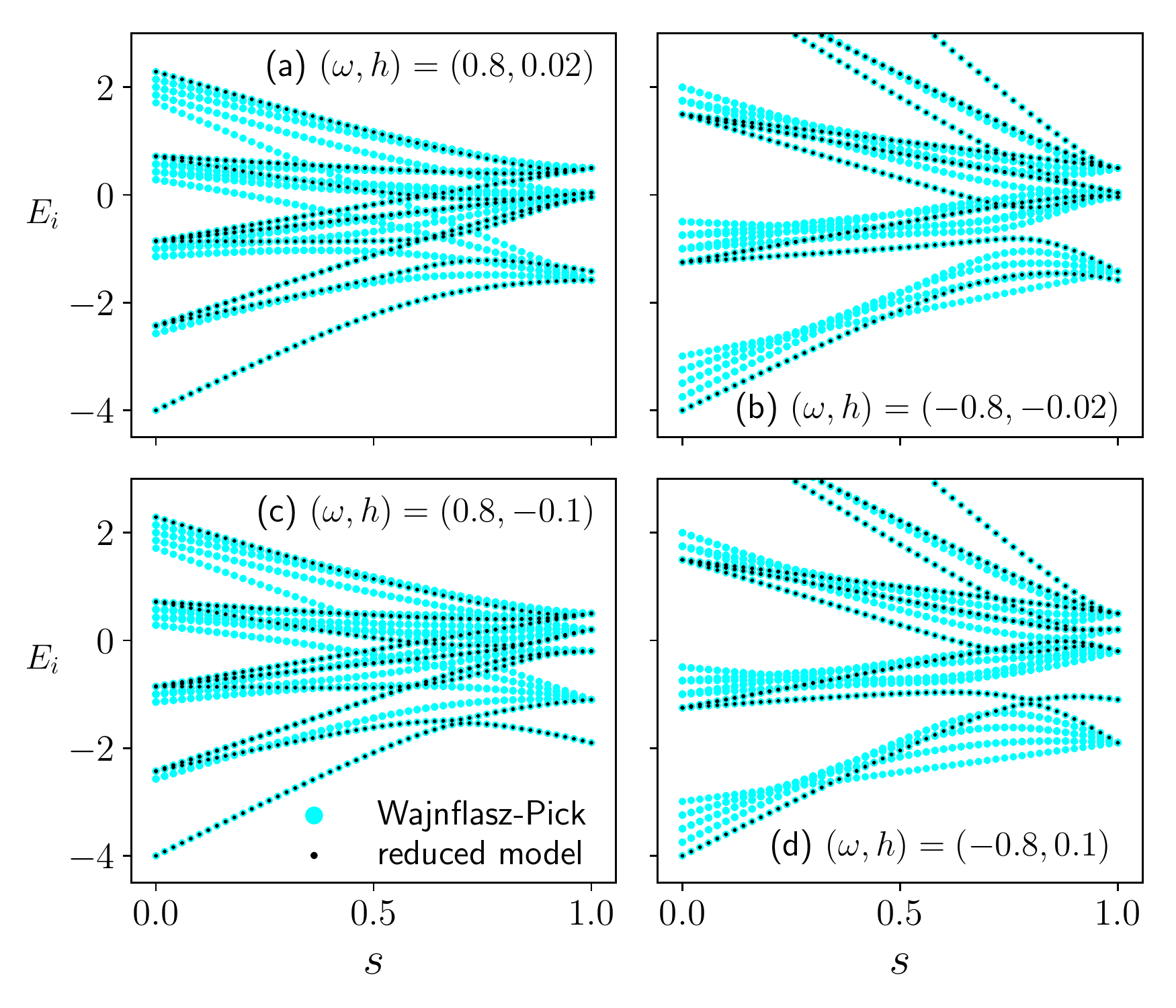}
\end{center}
\caption{Eigenenergies of the instantaneous Hamiltonian in the quantum Wajnflasz--Pick model (blue) and those in the reduced spin-1/2 model (black) as a function of $s \equiv t/T$. The parameters are the same as those in Fig.~\ref{fig2.fig}. 
}
\label{fig5.fig}
\end{figure}

Eigenvalue spectrum of the instantaneous Hamiltonian may help to understand these higher or lower success probabilities of the quantum Wajnflasz--Pick model than that of spin-1/2 model, although eigenvalues of the instantaneous Hamiltonian shows tangled spaghetti structures (Fig.~\ref{fig5.fig}). 
For example, in the case where $(\omega, h) = (0.8, -0.1)$, the energy gap between the ground state and the first excited state clearly closes once, which causes the low success probability (Panel (c) in Fig.~\ref{fig5.fig}). 
In the case where $(\omega, h) = (0.8, 0.02)$, the ground state and the first excited state are finally merged at the annealing time, where the degeneracy would cause the high success probability (Panel (a) in Fig.~\ref{fig5.fig}). 
However, according to the following discussion, it will be found that the latter explanation would not be correct in the case where $(\omega, h) = (0.8, 0.02)$. 
From panels (b) and (d) in Fig.~\ref{fig5.fig}, many crossings of eigenvalues are found to emerge. It suggests that there are no matrix elements in some states, and we may find symmetry behind the present quantum Wajnflasz--Pick model, where the Hamiltonian would be block diagonalized by a unitary operator $\hat U$. Since the energy spectrum of the original quantum Wajnflasz-Pick model shows very complicated behavior, it would be better to find out the reason of the high/low success probability from the reduced Hamiltonian, which are truly relevant for the efficiency of the quantum annealing.

For example, in the case where $(g_{\rm u}, g_{\rm l}) = (2,1)$, 
the single-spin Hamiltonian in the quantum Wajnflasz--Pick model is decomposable, where the irreducible representation is given by 
\begin{align}
	\hat U^{-1} \hat H(s) \hat U = 
	\begin{pmatrix}
		- h^{+} (s) & 0 & - 2\sqrt{2} h' (s)
		\\
		0 & - h^{-} (s) & 0 
		\\ 
		- 2\sqrt{2} h' (s) & 0 & h^z s
			\end{pmatrix}	, 
			\label{eq10}
\end{align} 
for arbitrary values of $s$, 
by using the unitary operator 
\begin{align}
	\hat U = \begin{pmatrix}
		\frac{1}{\sqrt{2}} & \frac{1}{\sqrt{2}} & 0 
		\\ 
		\frac{1}{\sqrt{2}} & - \frac{1}{\sqrt{2}} & 0
		\\ 
		0 & 0 & 1
	\end{pmatrix}, 
\end{align}
where $h^{\pm} (s) \equiv h^z s \pm 2 \omega h' (s)$, and $h' (s) \equiv (1-s)h^x/(2c)$. 
As a result, we may reduce a quantum annealing problem in the single-spin quantum Wajnflasz--Pick model into that of the spin-1/2 model, the Hamiltonian of which is given in the form 
\begin{align}
	\hat {\mathcal H} (s) = - [h^z s + \omega h'(s) ] \hat \sigma^z - 2 \sqrt{2} h'(s) \hat \sigma^x - \omega h'(s). 
\end{align}
Since the initial ground state of the single-spin Hamiltonian is given by $| \Psi (s=0) \rangle \propto  (c/2, c/2, 1)^{\rm T}$ in the original quantum Wajnflasz--Pick model, 
this state can be mapped to $\hat U |\Psi (s=0) \rangle \propto (c/\sqrt{2}, 0, 1)^{\rm T}$. 
It indicates that the initial ground state $\hat U |\Psi (s=0) \rangle $ can be also projected to the Hilbert space of the reduced Hamiltonian $\hat {\mathcal H} (s)$. 

\begin{figure}
\begin{center}
\includegraphics[width=8cm]{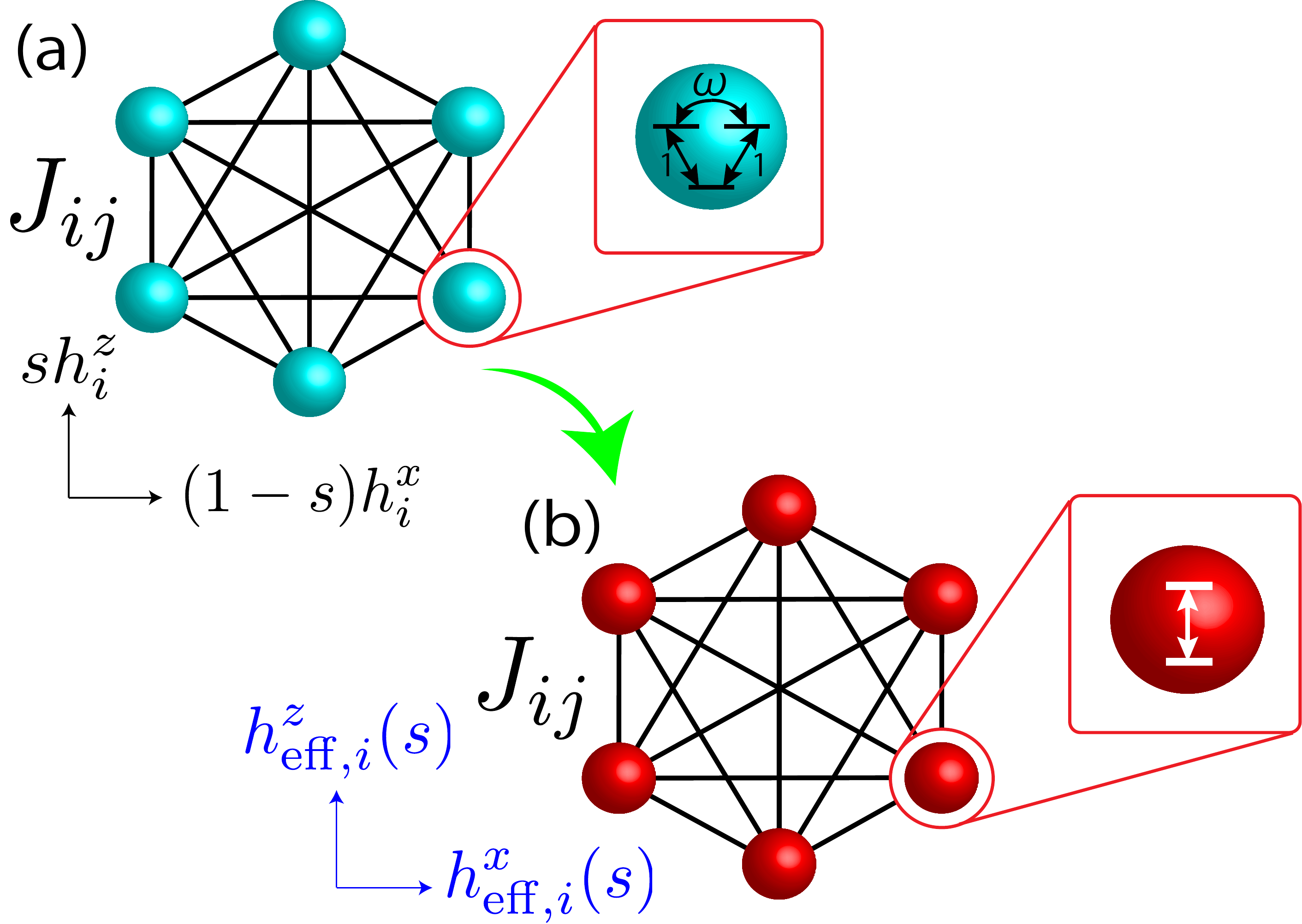}
\end{center}
\caption{Schematics of an original quantum Wajnflasz--Pick model (a) and its reduced model (b). }
\label{fig6.fig}
\end{figure}

This reduction of the single-spin problem in the case where $(g_{\rm u}, g_{\rm l}) = (2,1)$ can be generalized to an interacting $N$-spin problem (Fig.~\ref{fig6.fig}). A quantum annealing problem of the original quantum Wajnflasz--Pick model is reduced into that of the spin-1/2 model, given in the form 
\begin{align}
	\hat {\mathcal H} (s) = & s \left ( - \sum\limits_{i<j} J_{ij} \sigma_i^z \sigma_j^z \right ) 
	- \sum\limits_{i} h_{{\rm eff}, i}^z (s) \sigma_i^z 	
	\nonumber 
	\\ 
	& 
	- \sum\limits_{i} h_{{\rm eff}, i}^x (s) \sigma_i^x
	- \sum\limits_i \omega h_{i}' (s), 
	\label{EffectiveHamiltonian}
\end{align}
where 
\begin{align}
	h_{{\rm eff}, i}^z (s) \equiv & h_i^z s + \omega h'_i (s) , 
	\label{heffiz}
	\\ 
	h_{{\rm eff}, i}^x (s) \equiv & 2\sqrt{2} h_{i}' (s) , 
\end{align}
with
\begin{align}
	h_{i}' (s)	\equiv & \frac{(1-s) h_i^x}{2 c}. 
\end{align}

As in the single-spin case, the initial ground state of the original $N$-spin quantum Wajnflasz--Pick model can be also projected to the Hilbert space of the reduced Hamiltonian (\ref{EffectiveHamiltonian}). 
The coupling $J_{ij}$ in the reduced Hamiltonian is the same as that of the original Wajnflasz--Pick model. 
The effective longitudinal magnetic field $h_{{\rm eff}, i}^z$ in the reduced Hamiltonian also reaches the same value as that of the original Wajnflasz--Pick model at the end of the annealing: $h_{{\rm eff}, i}^z (s=1) = h_i^z$. 
Eigenvalues of the reduced spin-1/2 model exactly trace eigenvalues in the original Wajnflasz--Pick model (Fig.~\ref{fig5.fig}). 
The time-dependence of the ground state population of the problem Hamiltonian is confirmed to show the completely same behavior between the reduced model and the original model.

This effective model clearly explains behavior of success probability of the quantum Wajnflasz--Pick model shown in Fig.~\ref{fig3.fig}. 
Note that the coefficient $c$ is a positive real number such that the maximum eigenvalue of $\tau^x$ is unity, and we take $h_i^x = 1$. 
Then, $h_i'(s) \geq 0$ always holds during the annealing time $0 \leq s \leq 1$. 
In the case where the longitudinal magnetic field $h_i^z$ is very large, $|h_i^z | \gg |\omega| h_i' (0)$, 
the effect of the original longitudinal magnetic field $h_i^z$ is dominant compared with the effective additional term $\omega h'_i (s)$ except at the very early stage of the annealing $s \ll |\omega h_i'(0)/h_i^z|$. 
In this case, the problem Hamiltonian in the reduced model is almost the same as that in the conventional spin-1/2 model in Eq.~\eqref{eq2_0}. As a result, the success probability of the quantum Wajnflasz--Pick model is almost the same as that of the conventional spin-1/2 model, which provides $P \simeq P_{1/2}$. 

\begin{figure}
\begin{center}
\includegraphics[width=8cm]{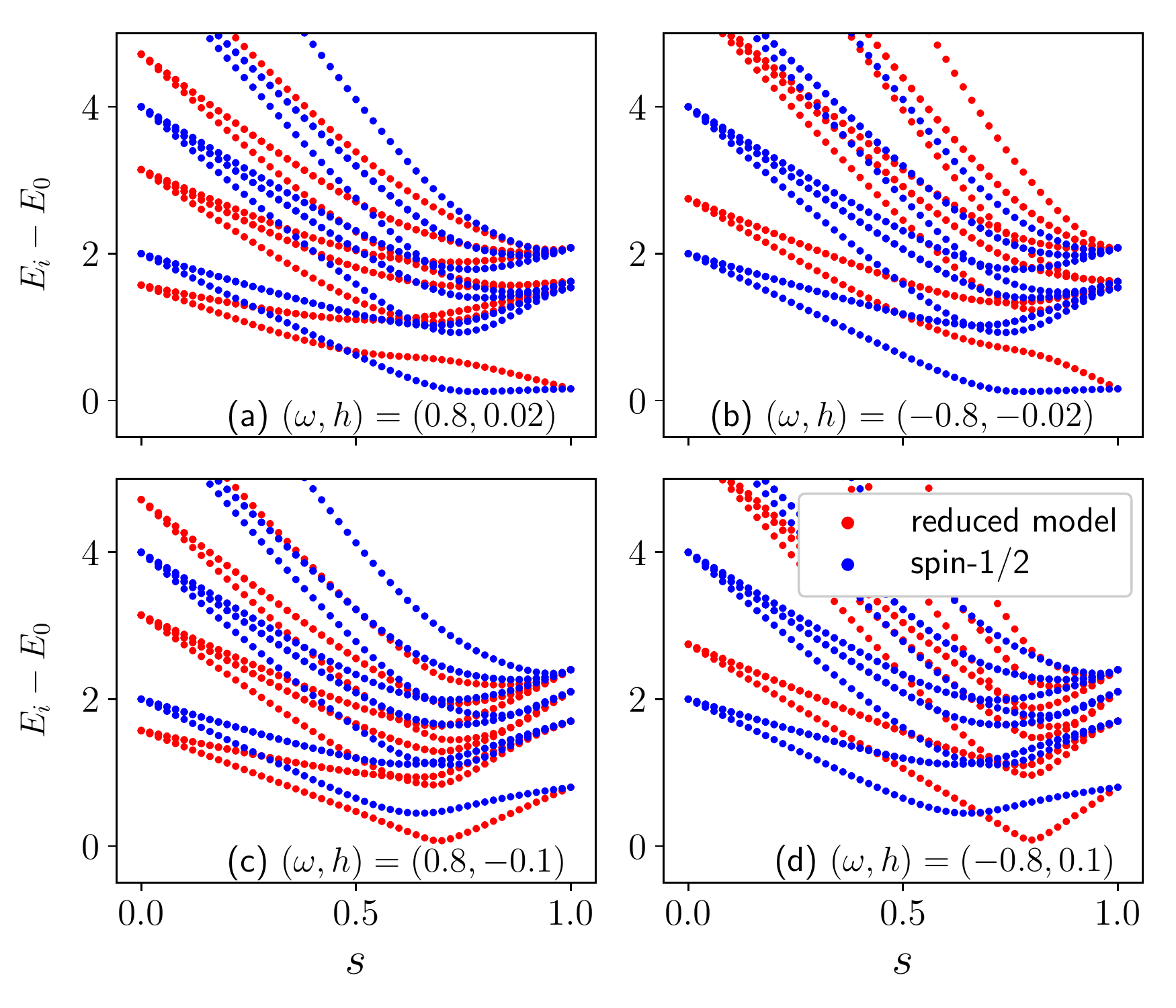}
\end{center}
\caption{Excited state energies measured from the ground state energy of the instantaneous Hamiltonian in the reduced model (red) and those in the conventional spin-1/2 model (blue). The parameters are the same as those in Fig.~\ref{fig2.fig}. 
}
\label{fig7.fig}
\end{figure}

In the case where the original longitudinal magnetic field $h_i^z$ is not large, the effective additional longitudinal magnetic field $\omega h_i'(s)$ cannot be neglected compared with $h_i^z$. 
When the effective additional field is in the same direction as the original longitudinal field, 
the total effective longitudinal magnetic field $h_{{\rm eff}, i}^z (s)$ is enhanced, which opens the energy gap between the ground state and the first excited state (Panels (a) and (b) in Fig.~\ref{fig7.fig}). 
This region is given by the condition $\omega h_i^z > 0$, which is consistent with the result shown in Fig.~\ref{fig3.fig}. 
As a result, the success probability of the quantum Wajnflasz--Pick model become superior to that of the conventional spin-1/2 model. 
When the effective additional field is in the opposite direction to the original longitudinal field, 
the total effective longitudinal magnetic field $h_{{\rm eff}, i}^z (s)$ is diminished, which closes the energy gap between the ground state and the first excited state (Panels (c) and (d) in Fig.~\ref{fig7.fig}). 
This region is given by the condition $\omega h_i^z < 0$, which is consistent with the result shown in Fig.~\ref{fig3.fig}. 
As a result, the success probability of the quantum Wajnflasz--Pick model become inferior to that of the conventional spin-1/2 model. 

\begin{figure}
\begin{center}
\includegraphics[width=8cm]{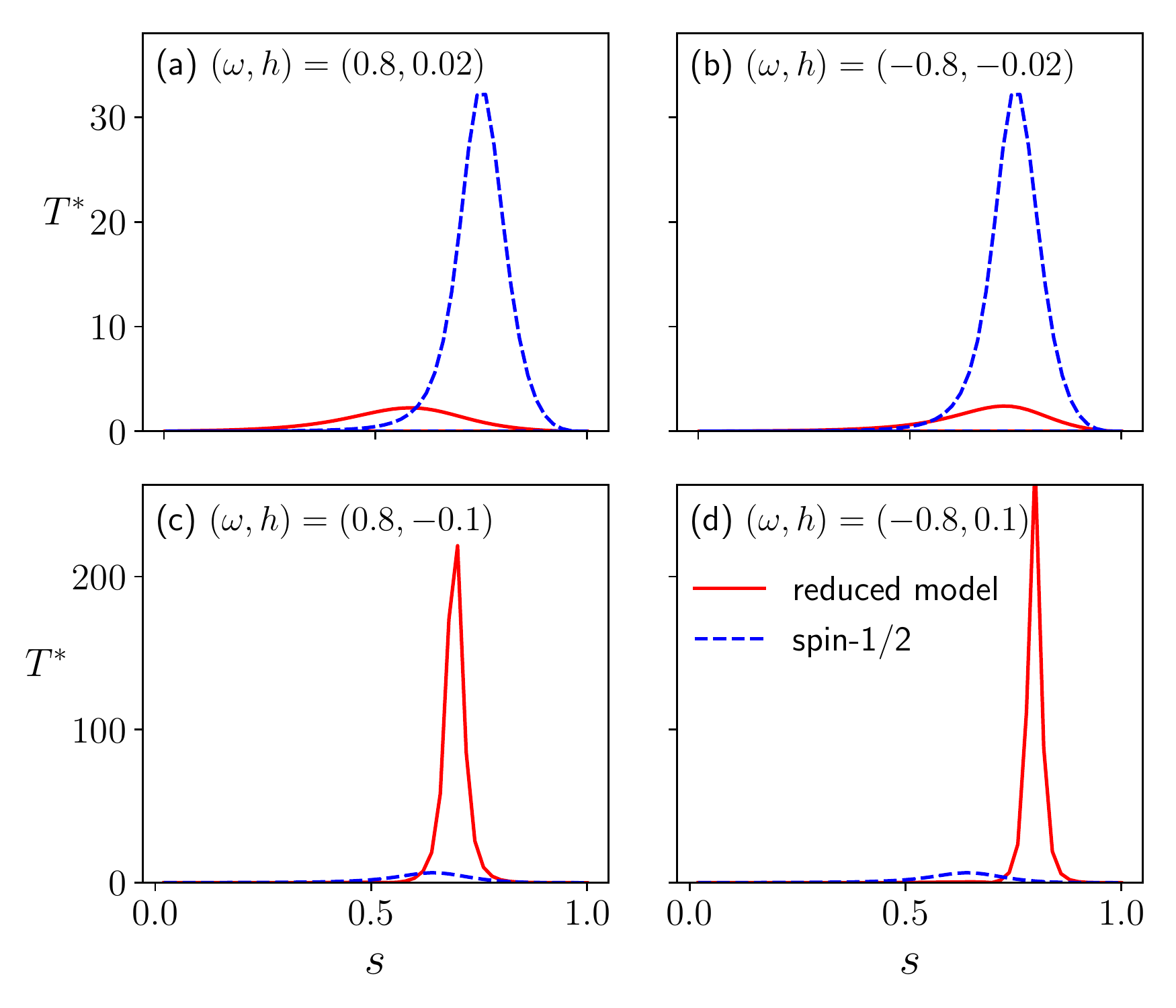}
\end{center}
\caption{Instantaneous reference annealing time $T^* \equiv b(s)/\Delta^2 (s)$ as a function of the scaled time $s$. 
The parameters are the same as those in Fig.~\ref{fig2.fig}. }
\label{fig8.fig}
\end{figure}

Behavior of success probability is also explained by the reference of the annealing time~\cite{Bapst2013}
\begin{align}
	\mathcal{T} \equiv  {\rm max}_s \left [ \frac{b(s)}{\Delta (s)^2} \right ], 
\end{align}
where
\begin{align}
	b(s) \equiv & \left | \langle \Psi_1 (s) | \frac{d \hat H (s)}{d s }  | \Psi_0 (s) \rangle \right | , 
	\\ 
	\Delta (s) \equiv & E_1 (s) - E_0 (s). 
\end{align}
Here, $| \Psi_{0(1)} (s) \rangle$ and $E_{0(1)} (s)$ are the wave functions and eigenenergies of the ground (first-excited) state with respect to the instantaneous Hamiltonian, respectively. 
Annealing machine needs the annealing time $T$ much larger than $\mathcal T$. 
Let $T^* \equiv b(s) /\Delta^2(s)$ be an instantaneous reference time of the annealing. 
The maximum value of this time $T^*$ in the reduced Wajnflasz--Pick model given in (\ref{EffectiveHamiltonian}) is suppressed compared with that of the conventional spin-1/2 model, 
where the effective additional field $\omega h'_i (s)$ is in the same direction as the original longitudinal field $h_i^z$ (Panels (a) and (b) in Fig.~\ref{fig8.fig}). 
It is consistent with the case where the quantum Wajnflasz--Pick model is more efficient than the conventional spin-1/2 model in the region where $\omega h_i^z > 0$ (Fig.~\ref{fig3.fig}). 
The maximum value of $T^*$ in the effective Wajnflasz--Pick model has larger values than that of the spin-1/2 model, 
where the effective additional field $\omega h'_i (s)$ is in the opposite direction to the original longitudinal field $h_i^z$ (Panels (c) and (d) in Fig.~\ref{fig8.fig}). 
It is consistent with the case where the quantum Wajnflasz--Pick model is less efficient than the conventional spin-1/2 model in the region where $\omega h_i^z < 0$ (Fig.~\ref{fig3.fig}).

\begin{figure}
\begin{center}
\includegraphics[width=8cm]{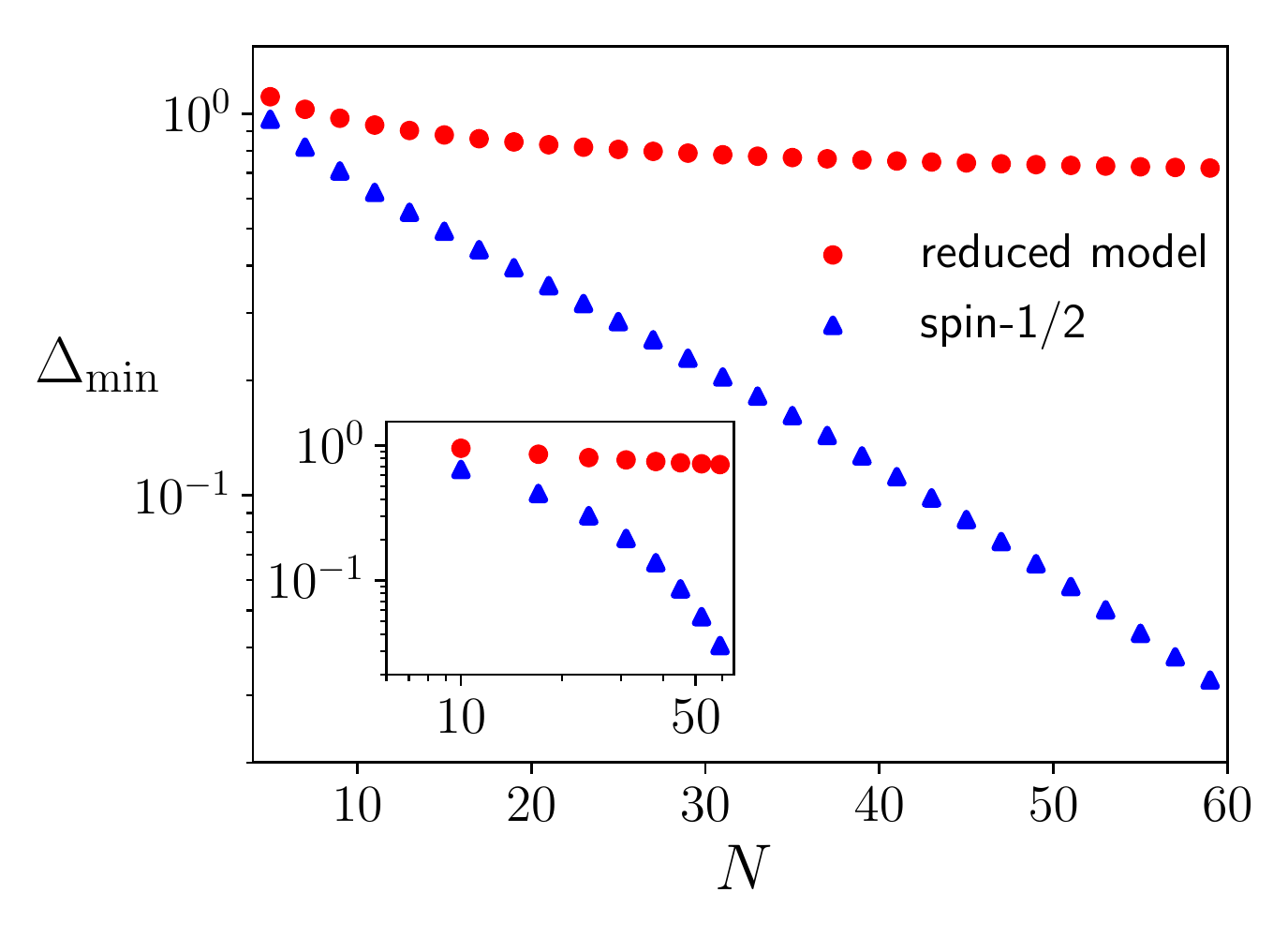}
\end{center}
\caption{Minimum energy gap as a function of the number of spins $N$ on a linear-log scale (a log-log scale in the inset). The minimum energy gap is obtained from the exact diagonalization of the ferromagnetic $p$-spin model with $p=3$. We have used $\omega = 0.8$ and $h^x = 1$. }
\label{fig9.fig}
\end{figure}

In order to perform the scaling analysis of the minimum energy gap $\Delta_{\rm min} \equiv {\rm min} [E_1  (s) - E_0(s) ]$, we consider the $p$-spin model in the absence of the longitudinal magnetic field: 
\begin{align}
\hat H(s) = s \left ( - \frac{1}{N^{p-1}}  \sum\limits_{i_1, \cdots, i_p }^N \hat \tau_{i_1}^z  \cdots  \hat \tau_{i_p}^z \right ) + (1-s) \left ( -h^x \sum\limits_i^N  \hat \tau_i^x \right ), 
	\label{pspinH}
\end{align}
where the transverse magnetic field is homogeneous. 
Replacement of $\tau_{i}^{x,y}$ with $\sigma_i^{x,y}$ provides the conventional $p$-spin model, where the first order phase transition emerges, and the minimum energy gap is known to close exponentially as $N$ increases in the case where $p$ is odd~\cite{Jorg2010A}. 
After mapping to the subspace spanned by the spin-$1/2$ model, the reduced Hamiltonian of the quantum Wajnflasz--Pick model with $(g_{\rm u}, g_{\rm l}) = (2,1)$ can be reduced to 
\begin{align}
\hat {\mathcal H} (s) = & - s\frac{1}{N^{p-1}}  \sum\limits_{i_1, \cdots, i_p }^N \hat \sigma_{i_1}^z  \cdots  \hat \sigma_{i_p}^z 
- (1-s) \Gamma^z \sum\limits_i^N  \hat \sigma_i^z 
\nonumber 
\\
& - (1-s) \Gamma^x \sum\limits_i^N  \hat \sigma_i^x , 
\\ 
= &  -s  \frac{1}{N^{p-1}} ( \hat M^z )^p 
- (1-s)   \Gamma^z  \hat M^z 
- (1-s)  \Gamma^x \hat M^x , 
	\label{pspinHreduced}
\end{align}
up to the constant energy shift, where $\Gamma^z \equiv \omega h^x / (2 c)$, $\Gamma^x \equiv \sqrt{2} h^x / c $, and $\hat M^{z, x} \equiv \sum\limits_i^N \hat \sigma_i^{z,x}$. 
By using the commutation relation $[\hat \sigma_i^x, \hat \sigma_j^z] = 2 i \hat \sigma_i^z \delta_{ij}$ and by following the standard argument of the angular momentum, where the total spin $\hat {\bf M}^2 \equiv (\hat M^x)^2 +  (\hat M^y)^2 +  (\hat M^z)^2$ conserves, 
the Hilbert space can be spanned by states $| J, M\rangle$, where $\hat {\bf M}^2 |J, M\rangle = J (J+2) | J, M \rangle$ and $\hat M^z | J, M \rangle = M | J, M \rangle$ with $M = -J, -J + 2, \cdots, J -2 , J$. The diagonal elements of this Hamiltonian is given by ${\mathcal H}_{MM} = - sM^p/(N^{p-1}) - (1-s)\Gamma^z M$, and the off-diagonal elements are ${\mathcal H}_{M,M\pm 2} = - (1-s)\Gamma^x \sqrt{J(J+2) - M (M\pm 2)}/2$. Since the ground state of this model is given by the case $J = N$, we diagonalize the $(N+1) \times (N+1)$ matrix of the reduced Hamiltonian. 
We compare the minimum energy gap of this model reduced from the quantum Wajnflasz--Pick model with that of the conventional $p$-spin model composed of the spin-$1/2$ system (Eq.~\eqref{pspinHreduced} with $\Gamma^z = 0$ and $\Gamma^x = h^x$). 
Figure~\ref{fig9.fig} clearly shows that the minimum energy gap closes exponentially in the conventional spin-$1/2$ model, and the gap closes polynomially in the model reduced from the quantum Wajnflasz--Pick model. This polynomial gap closing originates from the emergence of the effective longitudinal magnetic field in the reduced model: $\Gamma^z = \omega h^x / (2c) \neq 0$.

\section{Random coupling}

\begin{figure}
\begin{center}
\includegraphics[width=7cm]{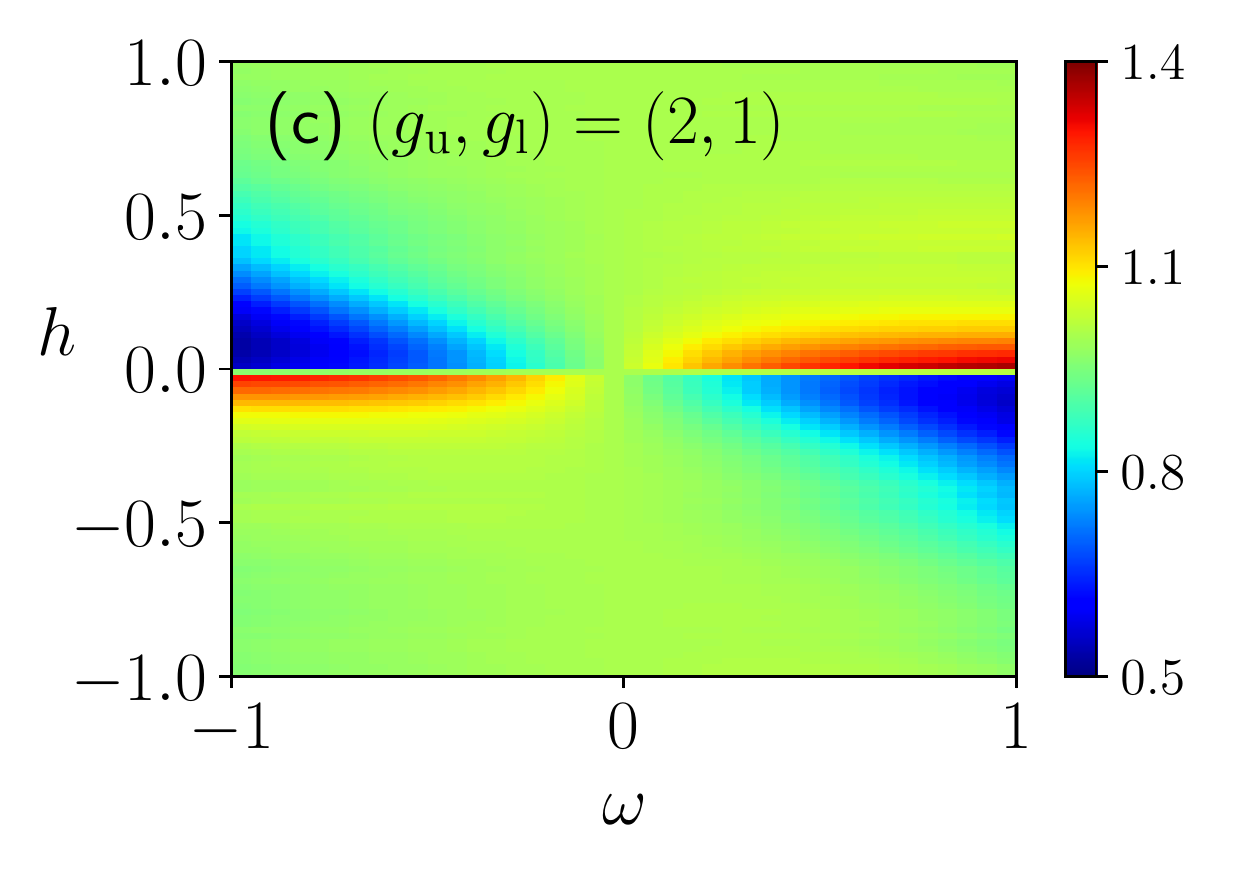}
\end{center}
\caption{Averaged success probability of a quantum Wajnflasz--Pick model $P$ scaled by that of spin-1/2 model $P_{1/2}$ in a randomly generated coupling strength case. 
We employed the coupling strength $J_{ij}$ randomly generated from the gaussian distribution function, where the mean is zero and the variance is $1/N$. 
We used $N=4$, $h_i^x = 1$ and $T=10$. 
The success probabilities $P$ and $P_{1/2}$ are averaged values of 100 samplings in each data point. } 
\label{fig10.fig}
\end{figure}

In the random spin-spin coupling case, where $J_{ij}$ are randomly generated by the gaussian distribution function~\cite{Gardner1985} 
\begin{align}
P(J_{ij}) = \sqrt{ \cfrac{N}{2 \pi } } \exp \left ( - \cfrac{N^{}}{2} J_{ij}^2 \right), 
\label{PJij}
\end{align}
the density plot of the mean-value of the success probability is similar to the uniform coupling case. 
The maximum (minimum) value of the success probability is, however, suppressed (increased) compared with the uniform coupling case (Fig.~\ref{fig10.fig}). 
The variances of the success probability of the quantum Wajnflasz--Pick model are almost ranged from 0.03 to 0.06 in the first and third orthants in the $\omega$-$h$ plane, where the higher success probability may be obtained than the conventional spin-1/2 model. 
They are almost ranged from 0.02 to 0.15 in the second and forth orthants in the $\omega$-$h$ plane, where the lower success probability may be obtained. 
In the spin-1/2 model, the variance of the success probability is almost within the range from 0.03 to 0.06 in all the orthants.

\begin{figure}
\begin{center}
\includegraphics[width=8.5cm]{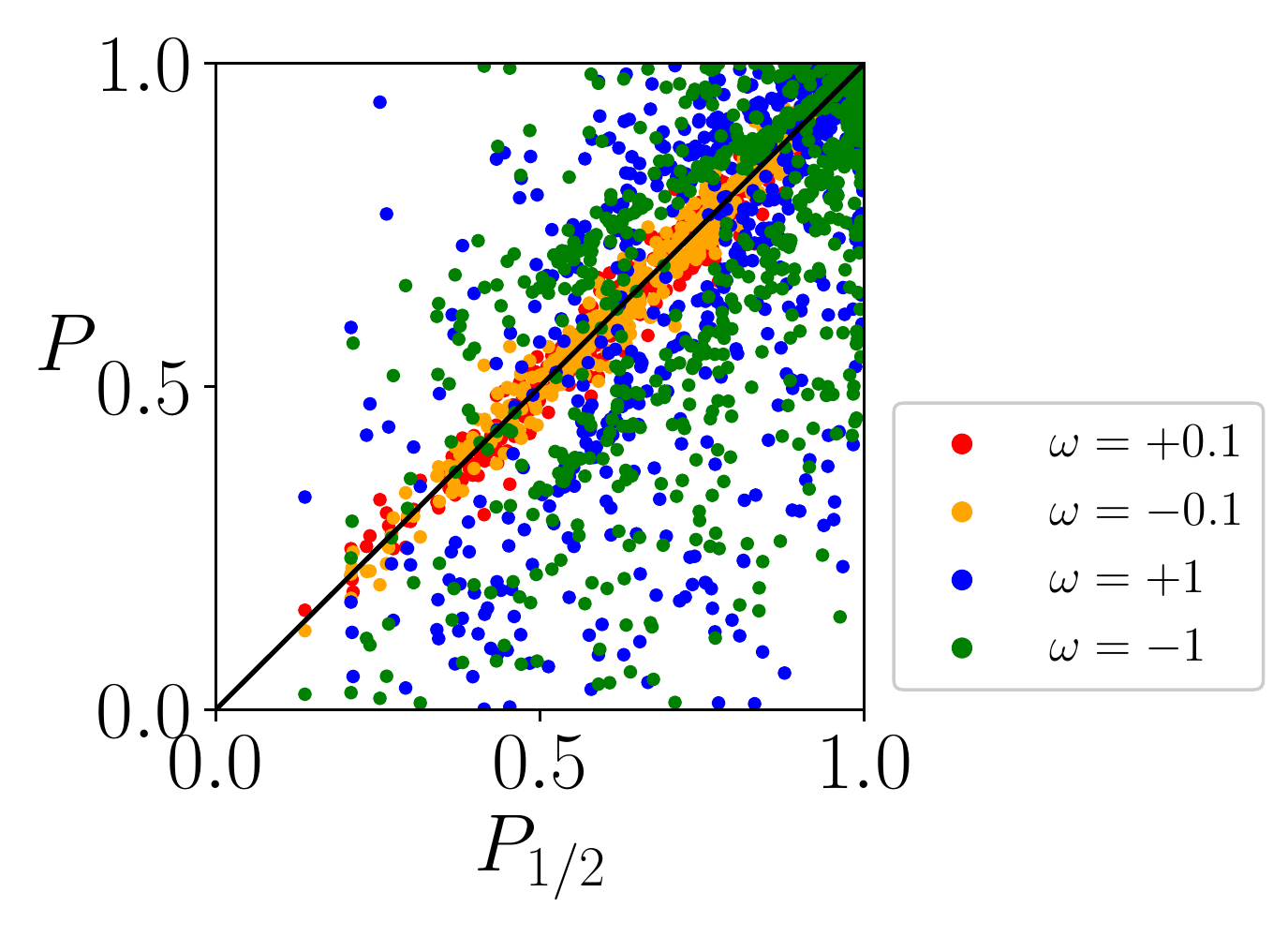}
\end{center}
\caption{Success probability $P$ of the quantum Wajnflasz--Pick model vs. success probability $P_{1/2}$ of the conventional spin-1/2 model. 
We take 1000 samples of problem hamiltonian with the random coupling strength $J_{ij}$ as well as the random longitudinal magnetic field $h_i^z$, both of which are generated from the standard Gaussian distribution. 
The means of $J_{ij}$ and $h_i^z$ are zeros, and the variances are $1/N$ and $1/\sqrt{2}$, respectively. 
For each problem set, we consider four cases $\omega = \pm 0.1$ and $\pm1$. 
We used $(g_{\rm u}, g_{\rm l}) = (2,1)$, $N=4$, $h_i^x = 1$ and $T=10$. 
}
\label{fig11.fig}
\end{figure}

The discussion above is in the case for a uniform longitudinal magnetic field. 
In the following, we discuss the case of random longitudinal magnetic fields $h_i^z$ in addition to the random interactions $J_{ij}$.   
The success probabilities $P$ and $P_{1/2}$ are almost equal in the weak internal state coupling case ($\omega = \pm 0.1$ in Fig.~\ref{fig11.fig}). 
In the strong internal state coupling case ($\omega = \pm 1$ in Fig.~\ref{fig11.fig}), the distribution is broaden. 
Although we can find cases where the conventional spin-1/2 model is superior to the quantum Wajnflasz--Pick model, 
we can also find many cases where the quantum Wajnflasz--Pick model is superior to the conventional spin-1/2 model, where the success probability is close to the unity compared with the conventional spin-1/2 model. 

In these random coupling cases, 
it may not be definitely concluded that the quantum Wajnflasz--Pick model is always more efficient than the conventional spin-1/2 model. 
The variance is relatively large, and there are cases where the quantum Wajnflasz--Pick model is inferior to the conventional spin-1/2 model (Fig.~\ref{fig11.fig}). 
However, we can find many cases where the quantum Wajnflasz--Pick model is possibly more efficient than the conventional spin-1/2 model. 
In the quantum Wajnflasz--Pick model and its reduced model, we have chances to find a better solution of the combinatorial optimization problem. 
In real annealing machines, we can extract a better solution after performing many sampling experiments by tuning $\omega$.

\section{Discussion}

In the case where $(g_{\rm u}, g_{\rm l}) = (2,1)$, the spin matrix in the quantum Wajnflasz--Pick model is represented by a $(3\times3)$-matrix, which suggests that the quantum Wajnflasz--Pick model in this case may be mapped into the model represented by the spin-1 matrices given by 
\begin{align}
\hspace{-10mm}
\hat S^x = \frac{1}{\sqrt{2}} \begin{pmatrix} 0 & 1 & 0 \\ 1 & 0 & 1 \\ 0 & 1 & 0 \end{pmatrix}, 
\hat S^y = \frac{i}{\sqrt{2}} \begin{pmatrix} 0 & -1 & 0 \\ 1 & 0 & -1 \\ 0 & 1 & 0 \end{pmatrix}, 
\hat S^z =                        \begin{pmatrix} 1 & 0 & 0 \\ 0 & 0 & 0 \\ 0 & 0 & -1 \end{pmatrix}. 
\end{align} 
Indeed, after we interchange elements of second and third rows in the spin matrices defined in Eq.~\eqref{matrixtauzx} in the case where $(g_{\rm u}, g_{\rm l}) = (2,1)$, as well as we interchange elements of second and third columns, simultaneously, we find the following maps  
\begin{align}
\hat \tau^z \mapsto & \,  \hat q^z \equiv \frac{2}{\sqrt{3}} \hat Q^{3z^2 -r^2} + \frac{1}{3},
\\ 
\hat \tau^x \mapsto & \, \hat q^x \equiv \cfrac{1}{c} [ \sqrt{2} \hat S^x + \Re\omega \hat Q^{x^2-y^2} - \Im\omega \hat Q^{xy}],
\end{align}
where we have introduced quadrupolar operators~\cite{Lauchli2006,Smerald2013}
\begin{align}
\hat Q^{3z^2 -r^2} \equiv & \frac{1}{\sqrt{3}} [2 (\hat S^z)^2 - (\hat S^x)^2 - (\hat S^y)^2], 
\\ 
\hat Q^{x^2-y^2}   \equiv & (\hat S^x)^2 - (\hat S^y)^2, 
\\ 
\hat Q^{xy}           \equiv & \hat S^x \hat S^y + \hat S^y \hat S^x , 
\end{align}
and $\Re\omega$ ($\Im \omega$) is the real (imaginary) part of $\omega$. 
Since $[\hat q^z, (\hat S^x)^2] = 0$ and $[\hat q^x, (\hat S^x)^2] = i (\Im \omega/c) \hat S^x$ hold, we find that $(\hat S^x)^2$ is the operator of the conserved quantity in the case where the parameter $\omega$ is a real number. 
The coupling of $\hat \tau_i^z \hat \tau_{j(\neq i)}^z$ is mapped into the interaction $\hat q_i^{z} \hat q_{j(\neq i)}^{z}$, which is a kind of the biquadratic interaction with respect to the spin. 
In short, the interacting quantum Wajnflasz--Pick model with $(g_{\rm u}, g_{\rm l}) = (2,1)$ can be mapped into the spin-1 model with an artificial biquadratic interaction. In particular, in the case where $\omega \in {\mathbb R}$, there is the hidden symmetry related to $(\hat S^x)^2$, which indicates that the quantum Wajnflasz--Pick model is reducible in this case.

It is general that an interacting quantum Wajnflasz--Pick model is reducible to the conventional spin-1/2 model. 
It holds for an arbitrary number of the degeneracy $(g_{\rm u}, g_{\rm l})$ and at an arbitrary time $s$, which can be proven in the case where the parameter $\omega$ is a real number and the condition $\omega > -1$ holds. 
In Supplementary Information, we show that the Hamiltonian of the interacting quantum Wajnflasz--Pick model with arbitrary $(g_{\rm u}, g_{\rm l})$ can be projected to the spin-1/2 model, and the initial ground state in the original quantum Wajnflasz--Pick Hamiltonian is also projected to the reduced Hilbert space. 
It indicates that the quantum annealing in the quantum Wajnflasz--Pick model can be always described by the reduced Hamiltonian. 

As shown in Supplementary Information, this projection holds not only in the $2$-body interacting quantum Wajnflasz--Pick model, 
but also in the $N$-body interacting model. 
It indicates that if the quantum Wajnflasz--Pick model is embedded into the Lechner--Hauke--Zoller (LHZ) architecture~\cite{Lechner2015,Glaetzle2017}, 
it can be also projected into the LHZ architecture composed of the spin-1/2 model, where the effective additional magnetic fields may emerge. 
The present quantum Wajnflasz--Pick model is a degenerate two-level system in the presence of the transverse magnetic field. 
The possibility of the implementation of  the degenerate two-level system has been discussed for the $D_2$ line of $^{87}$Rb~\cite{Margalit2013,Zhang2019}. 
The quantum Wajnflasz--Pick model is also similar to the $\Delta$-type cyclic artificial atom in the superconducting circuit~\cite{Liu2005,You2011}. 
In the $\Delta$-type artificial atom, the population is controllable by making use of the amplitudes and/or phases of microwave pulses, 
where the amplitudes alone controls the population in the conventional three-level system ($\Lambda$-type system)~\cite{Liu2005}. 
However, the $\Delta$-type system in the superconducting circuit is not an exactly degenerate two-level system. 
With this regard, it may be difficult to directly implement our model in the $\Delta$-type cyclic artificial atom in the superconducting circuit. 
Actually, it may be feasible to employ the spin-1/2 model with the scheduling function inspired by the quantum Wajnflasz--Pick model, in the case where the Schr\"odinger dynamics without the dissipation holds.

\begin{figure}
\begin{center}
\includegraphics[width=7cm]{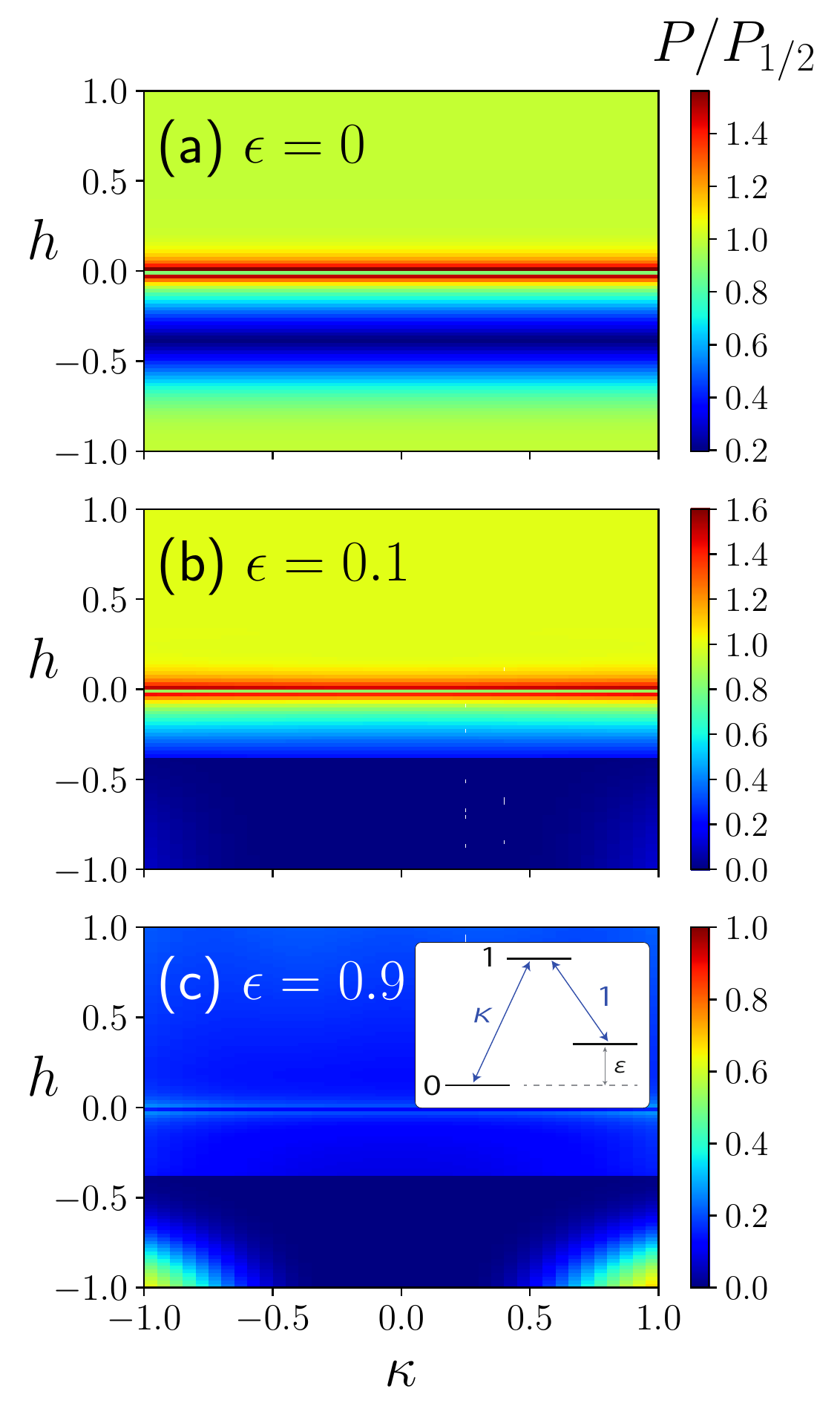}
\end{center}
\caption{
Success probability $P$ of the $\Lambda$-type system in the $h$-$\kappa$ plane, compared with that of the conventional spin-1/2 model $P_{1/2}$. 
The number of spin both in the $\Lambda$-type system and in the spin-1/2 model is $N=4$. We used the parameter sets $J_{ij} = 1/N$, $h_i^x = 1$ and $T=10$. 
}
\label{fig12.fig}
\end{figure}

The quantum Wajnflasz--Pick model is one of the qudit models, which is a kind of the artificial $\Delta$-type system~\cite{Liu2005,You2011} in the case where $(g_{\rm u}, g_{\rm l}) = (2,1)$. 
The question naturally arises whether the $\Lambda$-type system also shows the higher success probability than the conventional spin-1/2 model. 
The spin matrix of the $\Lambda$-type system we employ here is given by 
\begin{align}
\hat \tau^z = \begin{pmatrix} 0 & 0 & 0 \\ 0 & 1 & 0 \\ 0 & 0 & \varepsilon \end{pmatrix}, 
\hat \tau^x = \frac{1}{c} \begin{pmatrix} 0 & \kappa & 0 \\ \kappa & 0 & 1 \\ 0 & 1 & 0 \end{pmatrix}, 
\label{LambdaSpin} 
\end{align}
where we take $|\varepsilon| \leq 1$, and the coefficient $c \equiv \sqrt{1+\kappa^2}$ is a normalization factor so as the maximum eigenvalues of $\hat \tau^{x,z}$ are unity. 
The Hamiltonian of the quantum annealing with the $\Lambda$-type system is given by Eqs. (\ref{eq1}), (\ref{eq2}), and (\ref{eq3}), where $\hat \tau^{x,z}$ are replaced with those given in (\ref{LambdaSpin}). 
The success probability in the $\Lambda$-type system is found to be higher than that in the conventional spin-1/2 model, in the case where $\varepsilon$ is small in the weak longitudinal magnetic field region, which is similar to the case of the quantum Wajnflasz--Pick model (Panels (a) and (b) in Fig.~\ref{fig12.fig}). 
When $\epsilon$ is large, on the other hand, the success probability is drastically suppressed (Panel (c) in Fig.~\ref{fig12.fig}). 
In the case of a single $\Lambda$-spin system with $\epsilon = 0$, which corresponds to a degenerate two-level system, 
the unitary transformation 
\begin{align}
	\hat U = \frac{1}{c} \begin{pmatrix}
		\kappa & 0 & 1 \\ 0 & 1 & 0 \\ 1 & 0 & -\kappa
	\end{pmatrix}
\end{align} 
can map the Hamiltonian $\hat H (s) = - s h^z \hat \tau^z - (1-s) h^x \hat \tau^x$ to the following block diagonal form: 
\begin{align}
	\hat U^{-1} \hat H (s) \hat U = \begin{pmatrix}
		0 & - (1-s) h^x & 0
		\\ 
		 - (1-s) h^x & - s h^z & 0
		\\ 
		0 & 0 & 0
	\end{pmatrix}. 
\end{align}
As a result, after exchanging the first and second columns and also the first and second rows, we may reduce a quantum annealing problem in this $\Lambda$-spin model into that of the spin-1/2 model, the Hamiltonian of which is given by $\hat {\mathcal H} (s) = - s h^z \sigma^z/2 - (1-s)h^x \sigma^x - sh^z/2$. 
Although the $\Lambda$-type system may provide the higher success probability than the conventional spin-1/2 model, the effect of dark states (never employed states) on the quantum annealing in the general $\Lambda$-spin case and its reduction to the spin-1/2 model in the many-spin system would be important issues for future study. 

To summarize, we have demonstrated that qudit models, such as the quantum Wajnflasz--Pick model as well as the $\Lambda$-type system, may provide the higher success probability than the conventional spin-1/2 model in the weak magnetic field region. 
We have analytically shown that the quantum Wajnflasz--Pick model can be reduced into the spin-1/2 model, where effect of the transverse magnetic field in the original Hamiltonian emerges as the effective additional longitudinal magnetic field in the reduced Hamiltonian, which possibly opens the energy gap between the ground state and the first excited state in the reduced Hamiltonian. 
Since qubits have experimental advantages for the manipulation, the direct implementation of the reduced spin-1/2 model may be convenient for the quantum annealing. 
On the other hand, the reduction to the subspace in terms of the spin-1/2 model is useful only in the case where we focus on the Schr\"odinger dynamics. If we consider the dissipation as a realistic system, the transition between the subspaces emerges. The efficiency of the quantum annealing in this system is open for further study.

\section{Conclusions}
We studied the performance of the quantum annealing constructed by one of the degenerate two-level systems, called the quantum Wajnflasz--Pick model. 
This model shows the higher success probability than the conventional spin-1/2 model in the region where the longitudinal magnetic field is weak. 
The physics behind this is that the quantum annealing of this model can be reduced into that of the spin-1/2 model, 
where the effective longitudinal magnetic field in the reduced Hamiltonian may open the energy gap between the ground state and the first excited state, which gives rise to the suppression of the Landau--Zener transition. 
The reduction of the quantum Wajnflasz--Pick model to the spin-1/2 model is general at an arbitrary time as well as in an arbitrary number of degeneracies. 
We also demonstrated that the $\Lambda$-type system also shows the higher success probability than the conventional spin-1/2 model in the weak magnetic field regions. 
We hope that studying quantum annealing with variant spins, and utilizing the insight of their reduced model will promote further development of high performance quantum annealer. 

\section*{Acknowledgement}
We thank R. van Bijnen, W. Lechner, Y. Matsuzaki, T. Ishikawa, T. Yamamoto, and T. Nikuni for fruitful discussions and comments. 
Two of the authors (S.W. and S.K.) were supported by Nanotech CUPAL, Japan Science and Technology Agency (JST). 
Y.S. and S.K. were supported by the New Energy and Industrial Technology Development Organization (NEDO), Japan. 

%
%

\section*{Supplemental Information}\label{sec:AppendixA}

We have shown in \eqref{EffectiveHamiltonian} that the quantum Wajnflasz--Pick model with the $N$ spins with $(g_{\rm u} , g_{\rm l}) = (2,1)$ can be reduced into the spin-1/2 model. In the single-spin case, in particular, 
if we take 
\begin{align}
    \ket{\phi_{\text{u}}}_{} &= \frac{1}{\sqrt{2}} (1, 1, 0)^{\rm T}, \\
    \label{eq:basis l}
    \ket{\phi_{\text{l}}}_{} &= (0,0,1)^{\rm T}, 
\end{align} 
the Hamiltonian \eqref{eq1} projected into the Hilbert space spanned by $\ket{\phi_{\text{u,l}}}_{}$ is represented by 
\begin{align}
    \hat {\mathcal H} (s) = & 
    \begin{pmatrix}
    	\bra{\phi_{\rm u}} \hat H(s) \ket{\phi_{\rm u}} & \bra{\phi_{\rm u}} \hat H(s) \ket{\phi_{\rm l}}
	\\ 
    	\bra{\phi_{\rm l}} \hat H(s) \ket{\phi_{\rm u}} & \bra{\phi_{\rm l}} \hat H(s) \ket{\phi_{\rm l}}	
    \end{pmatrix}
    \\ 
    = & 
	\begin{pmatrix}
		- h^z s - 2 \omega h' (s)  & - 2\sqrt{2} h' (s)
		\\ 
		- 2\sqrt{2} h' (s)  & h^z s
	\end{pmatrix}. 
\end{align} 
It can be clearly found that this representation is exactly the same form as the block matrix shown in \eqref{eq10}.

In this Supplementary Information, by generalizing this idea, we show that the quantum annealing problem of the quantum Wajnflasz--Pick model with the $N$-spins with arbitrary number of degeneracy $g_{\rm u,l}$ can be reduced into the spin-1/2 model. 
First, consider the Hamiltonian of the quantum Wajnflasz--Pick model 
\begin{align}
    \hat{H}(s) &= s \hat{H}_z + (1-s) \hat{H}_x, 
    \label{eq:Hamiltonian of QWP}
\end{align}
where $\hat{H}_x \equiv  - \sum_{i=1}^{N} h_i^x \hat{\tau}_{i}^{x}$ with $h_i^x > 0$, and $\hat{H}_z \equiv f(\hat{\tau}_{1}^{z}, \dotsc ,\hat{\tau}_{N}^{z})$ can be expanded in the Maclaurin series. 
Equation \eqref{eq:Hamiltonian of QWP} is a generalization of (\ref{eq1}) with (\ref{eq2}) and (\ref{eq3}). 
We here introduce eigenstates of  a single spin $\hat \tau_i^z$ at a site $i$ as 
\begin{gather}
    \label{eq:upper states}
    \ket{u_{1}}_{i}, \ket{u_{2}}_{i}, \dotsc , \ket{u_{g_{\text{u}}}}_{i}, \\
    \label{eq:lower states}
    \ket{l_{1}}_{i}, \ket{l_{2}}_{i}, \dotsc , \ket{l_{g_{\text{l}}}}_{i},
\end{gather}
where $\ket{u_{k}}_{i}$ for $k = 1, 2, \cdots, g_{\rm u}$ and  $\ket{l_{k}}_{i}$ for $k = 1, 2, \cdots, g_{\rm l}$ are eigenstates whose eigenvalues of $\hat \tau_i^z$ is $+1$ and $-1$, respectively. 

In the following, we first prove a lemma I:  if the parameter $\omega$ is a real number, the Hamiltonian of the $N$-spin quantum Wajnflasz--Pick model can be decomposed into two parts 
\begin{align}
    \label{eq:block diag}
    \hat{H} = \hat{P}\hat{H}\hat{P} + (\hat{1} - \hat{P})\hat{H}(\hat{1} - \hat{P}), 
\end{align}
where $\hat{P} \equiv \bigotimes_{i=1}^{N}\hat{P}_{i}$ with $\hat{P}_{i} = \ket{\phi_{\text{u}}}_{i}\bra{\phi_{\text{u}}}_{i} + \ket{\phi_{\text{l}}}_{i}\bra{\phi_{\text{l}}}_{i}$ 
is a projection operator. A local projection operator $\hat{P}_{i}$ is spanned by two bases $ \ket{\phi_{\text{u,l}}}_{i}$, where 
\begin{align}
    \label{eq:basis u}
    \ket{\phi_{\text{u}}}_{i} &\equiv \frac{1}{\sqrt{g_{\text{u}}}}\sum_{k=1}^{g_{\text{u}}}\ket{u_{k}}_{i}, \\
    \label{eq:basis l}
    \ket{\phi_{\text{l}}}_{i} &\equiv \frac{1}{\sqrt{g_{\text{l}}}}\sum_{k=1}^{g_{\text{l}}}\ket{l_{k}}_{i}. 
\end{align} 
It indicates that the Hilbert space of the quantum Wajnflasz--Pick model can be reduced to a subspace spanned by $ \ket{\phi_{\text{u,l}}}_{i}$. 
We also prove a lemma II: if the parameter $\omega$ is a real number and the condition $\omega > -1$ holds, 
the ground state of the initial Hamiltonian $\hat{H}_{x}$ with $h_i^x > 0$ is an element of the Hilbert space spanned by $\ket{\phi_{\text{u,l}}}_{i}$. 
According to these two lemmas I and II, in the case where $\omega \in \mathbb{R}$ and $\omega > -1$, the quantum annealing problem in the quantum Wajnflasz--Pick model is represented as a model where the local spin has two states---the spin-1/2 model. 

We first consider the lemma I. Since $\hat P^2 = \hat P$, a necessary and sufficient condition providing \eqref{eq:block diag} is given by 
\begin{align}
    \label{eq:rel comm P H zero}
    \left[\hat{P}, \hat{H} \right] = 0. 
\end{align} 
The condition \eqref{eq:rel comm P H zero} can be reduced into 
\begin{align}
    \label{eq:comm rel N spin system}
    \left[ \hat{P}, \hat{H} \right]
    &=
    s \left[ \hat{P}, \hat{H}_{z} \right] + (1-s) \left[ \hat{P}, \hat{H}_{x} \right] = 0. 
\end{align}
Here, we will easily prove that $[ \hat{P}, \hat{H}_{z} ]  = 0$, for the projection operator $\hat P$ is composed of bases that diagonalize $\hat H_z$. 
Indeed, since $\hat H_z$ is composed of $\hat{\tau}_{i}^{z}$, what we need to show is 
\begin{align}
    \label{eq:sufficient condition z}
    [\hat{P}_{i}, \hat{\tau}_{i}^{z}]= 0. 
\end{align}
A spectral representation of the single-spin operator $\hat \tau_i^z$ can be represented by eigenstates of $\hat \tau_i^z$, given in the form 
\begin{align}
    \label{eq:tauz spectrum repr}
    \hat{\tau}_{i}^{z}
    &=
    \sum_{k=1}^{g_{\text{u}}} \ket{u_{k}}_{i}\bra{u_{k}}_{i}
    - \sum_{k=1}^{g_{\text{l}}} \ket{l_{k}}_{i}\bra{l_{k}}_{i}.
 \end{align}
The projection operator $\hat P_i$ is also composed of eigenstates of $\hat{\tau}_{i}^{z}$. 
We can thus immediately conclude that \eqref{eq:sufficient condition z} holds, which provides $[ \hat{P}, \hat{H}_{z} ]  = 0$. 
The result (\ref{eq:sufficient condition z}) can also provide the following representation 
\begin{align}
    \label{eq:block diag i z}
    \hat{\tau}_{i}^{z} &= \hat{P}_{i}\hat{\tau}_{i}^{z}\hat{P}_{i}
                         + (\hat{1}_{i} - \hat{P}_{i})\hat{\tau}_{i}^{z}(\hat{1}_{i} - \hat{P}_{i}). 
\end{align}

The remain we need to prove is 
\begin{align}
    \label{eq:comm rel in Re omega}
   \left[ \hat{P} , \hat{H}_{x} \right]
    &=
    - \sum_{j=1}^{N} h_j^x \left[ \bigotimes_{i=1}^{N}\hat{P}_{i}, \hat{\tau}_{j}^{x} \right] =0 . 
\end{align}
The necessary and sufficient condition for \eqref{eq:comm rel in Re omega} for arbitrary $h_j^x$ is 
\begin{align}
    \label{eq:sufficient condition i}
    [\hat{P}_{i}, \hat{\tau}_{i}^{x}] = 0, 
\end{align}
because we can expand a term in \eqref{eq:comm rel in Re omega} in the following way 
\begin{align}
    \label{eq:comm rel Hx and P 2}
    \left[ \bigotimes_{i=1}^{N}\hat{P}_{i}, \hat{\tau}_{j}^{x} \right]
    &=
    \left(\bigotimes_{i=1}^{j-1}\hat{P}_{i}\right)
    \left[\hat{P}_{j}, \hat{\tau}_{j}^{x} \right]
    \left(\bigotimes_{i=j+1}^{N}\hat{P}_{i}\right) . 
\end{align}
In order to show \eqref{eq:sufficient condition i}, it is convenient to introduce the spectral representation of the single-spin operator $\hat \tau^{x}_{i}$, given by 
\begin{align}
    \label{eq:taux spectrum repr}
    \hat{\tau}_{i}^{x}
    &=
    \frac{1}{c} ( \omega \hat A_i + \hat B_i + {\rm H.c.} ) ,  
\end{align}
where 
\begin{align}
\hat A_i \equiv &  \sum_{\substack{k,k'=1\\k > k'}}^{g_{\text{u}}} \ket{u_{k'}}_{i}\bra{u_{k}}_{i} + \sum_{\substack{k,k'=1\\k > k'}}^{g_{\text{l}}} \ket{l_{k'}}_{i}\bra{l_{k}}_{i}, 
\\ 
\hat B_i \equiv & \sum_{k=1}^{g_{\text{u}}}\sum_{k'=1}^{g_{\text{l}}} \ket{u_{k}}_{i}\bra{l_{k'}}_{i}. 
\end{align} 
Here, a constant $c$ is the normalization factor such that the spectral norm of $\hat{\tau}_{i}^{x}$ is to be unity. 
We can also represent $\hat{\tau}_{i}^{x}$ as 
\begin{align}
    \label{eq:taux effective spectrum repr}
    \hat{\tau}_{i}^{x}
    &=
    \frac{1}{c} \left [ 
        \frac{1}{2} \Re\omega \left(
            \hat A_i^{\rm u} + \hat A_i^{\rm l}
        \right)
        + i\Im \omega \hat A_i + \hat B_i' +  {\rm H.c.}
        \right ] 
\end{align}
where
\begin{align}
\hat A_i^{\rm u} \equiv & g_{\text{u}}\ket{\phi_{\text{u}}}_{i}\bra{\phi_{\text{u}}}_{i}
            - \sum_{k=1}^{g_{\text{u}}}\ket{u_{k}}_{i}\bra{u_{k}}_{i}, 
\\
\hat A_i^{\rm l} \equiv & g_{\text{l}}\ket{\phi_{\text{l}}}_{i}\bra{\phi_{\text{l}}}_{i}
            - \sum_{k=1}^{g_{\text{l}}}\ket{l_{k}}_{i}\bra{l_{k}}_{i} , 
\\ 
\hat B_i' \equiv &  \sqrt{g_{\text{u}}g_{\text{l}}}     \ket{\phi_{\text{u}}}_{i}\bra{\phi_{\text{l}}}_{i} .           
\end{align}
Here, $\Re\omega$ and $\Im\omega$ in (\ref{eq:taux effective spectrum repr}) are the real and imaginary parts of $\omega$, respectively. 
By using the representation \eqref{eq:taux effective spectrum repr}, we can obtain the following result: 
\begin{align}
    \label{eq:comm rel P i taux}
    \left[ \hat{P}_{i}, \hat{\tau}_{i}^{x} \right]
    &=
    [\ket{\phi_{\text{u}}}_{i}\bra{\phi_{\text{u}}}_{i}, \hat{\tau}_{i}^{x}]
    + [\ket{\phi_{\text{l}}}_{i}\bra{\phi_{\text{l}}}_{i}, \hat{\tau}_{i}^{x}]
    \notag \\
    &=
    i\Im \omega (\hat C_i^{\rm u} + \hat C_i^{\rm l} + {\rm H.c.}), 
\end{align}
where 
\begin{align}
\hat C_i^{\rm u} \equiv & 
    \frac{1}{c\sqrt{g_{u}}}
        \sum_{k=1}^{g_{\text{u}}}(2k- g_{u} - 1)
           \ket{\phi_{\text{u}}}_{i}\bra{u_{k}}_{i} , 
    \notag \\
\hat C_i^{\rm l} \equiv & 
    \frac{1}{c\sqrt{g_{l}}}
    \sum_{k=1}^{g_{\text{l}}}(2k- g_{l} - 1)
       \ket{\phi_{\text{l}}}_{i}\bra{l_{k}}_{i} . 
\end{align}
From the result \eqref{eq:comm rel P i taux}, we find that \eqref{eq:sufficient condition i} holds in the case where $\omega$ is a real number: $\Im \omega = 0$, 
which provides 
\begin{align}
    \label{eq:block diag i x}
    \hat{\tau}_{i}^{x} &= \hat{P}_{i}\hat{\tau}_{i}^{x}\hat{P}_{i}
                         + (\hat{1}_{i} - \hat{P}_{i})\hat{\tau}_{i}^{x}(\hat{1}_{i} - \hat{P}_{i}). 
\end{align}
As a result, \eqref{eq:rel comm P H zero} is found to be hold, and the Hamiltonian $\hat H$ are reducible and block diagonalizable, independent of the time $s$ as well as a structure of the spin coupling. 

Finally, we will prove the lemma II, where the ground state of the initial Hamiltonian $\hat{H}_{x}$ with $h_i^x > 0$ is an element of the Hilbert space spanned by $\ket{\phi_{\text{u,l}}}_{i}$, if the conditions $\omega \in \mathbb{R}$ and $\omega > -1$ hold. 
In the following, we assume that the parameter $\omega$ is a real number. 
Since the initial driver Hamiltonian $\hat H_x$ is a sum of a single-site spin operator $-h_i^x\hat{\tau}_{i}^{x}$ with $h_i^x > 0$, 
let $|\Psi (s=0) \rangle \equiv \bigotimes_{i=1}^{N}\ket{\psi_{0}}_{i}$ be the initial ground state, where $\ket{\psi_{0}}_{i}$ is the ground state of $- h_i^x \hat{\tau}_{i}^{x}$ with $h_i^x > 0$. 
In order to prove the lemma II, 
it is sufficient to show that the state $\ket{\psi_{0}}_{i}$ is an element of the Hilbert space spanned by $ \ket{\phi_{\text{u,l}}}_{i}$. 

We construct eigenstates of $- \hat{\tau}_{i}^{x}$, whose number is  ($g_{\rm u} + g_{\rm l}$), given in the form 
\begin{align}
    \label{eq:deg init gs u}
    \ket{m_{\text{u}}}_{i}
    &\equiv
    \frac{1}{\sqrt{g_{\text{u}}}}\sum_{k=1}^{g_{\text{u}}}
    e^{2\pi i(k-1)m_{\text{u}} / g_{\text{u}}} \ket{u_{k}}_{i}
    \\
    \label{eq:deg init gs l}
    \ket{m_{\text{l}}}_{i}
    &\equiv
    \frac{1}{\sqrt{g_{\text{l}}}}\sum_{k=1}^{g_{\text{l}}}
    e^{2\pi i(k-1)m_{\text{l}} / g_{\text{l}}} \ket{l_{k}}_{i}
    \\ 
    \label{eq:init gs}
    \ket{\lambda_{\pm}}_{i}
    &\equiv  \sqrt{g_{\text{u}}} a_{\pm} \ket{\phi_{\text{u}}}_{i}
                        +\sqrt{g_{\text{l}}} \ket{\phi_{\text{l}}}_{i}, 
\end{align}
where $m_{\text{u,l}} = 1, 2, \dotsc, g_{\text{u,l}}-1$. 
Eigenvalues of $-\hat{\tau}_{i}^{x}$ for $\ket{m_{\text{u,l}}}_{i}$ are given by $\omega/c$. 
The states $\ket{\lambda_{\pm}}_{i}$ become eigenstates of $-\hat{\tau}_{i}^{x}$, if we take 
\begin{align}
    \label{eq:def a}
    a_{\pm} \equiv \frac{(g_{\text{u}} - g_{\text{l}})\omega
                    \mp \sqrt{(g_{\text{u}} - g_{\text{l}})^{2}\omega^{2}
                              + 4g_{\text{u}}g_{\text{l}}}}{2g_{\text{u}}}, 
\end{align}
whose eigenvalues of $-\hat{\tau}_{i}^{x}$ are given by 
\begin{align}
    \label{eq:lambda}
    \lambda_{\pm} = \frac{1}{c}\left\{
        \left(1-\frac{g_{\text{u}}+g_{\text{u}}}{2}\right)\omega
        \pm \frac{1}{2}\sqrt{(g_{\text{u}} - g_{\text{l}})^{2}\omega^{2}
                              + 4g_{\text{u}}g_{\text{l}}}
    \right\}. 
\end{align}
These eigenstates $\ket{m_{\text{u,l}}}_{i}$ and $\ket{\lambda_{\pm}}_{i}$ are orthogonal. 
We can find that the relations $\lambda_{+} > \lambda_{-}$ as well as $(\omega/c) > \lambda_-$ hold in the case where $\omega > -1$. 
As a result, $h_i^x \lambda_{-}$ is the minimum eigenvalue of $- h_i^x \hat{\tau}_{i}^{x}$ with $h_i^x > 0$, and the ground state is $\ket{\lambda_-}_i = \sqrt{g_{\text{u}}} a_{-} \ket{\phi_{\text{u}}}_{i} +\sqrt{g_{\text{l}}} \ket{\phi_{\text{l}}}_{i}$, 
which indicates that the initial ground state is an element of the Hilbert space spanned by $\ket{\phi_{\text{u,l}}}_{i}$. 

To summarize, in the case where $\omega$ is a real number and $\omega > -1$ holds, 
the quantum annealing problem
 in the quantum Wajnflasz--Pick model can be exactly described by the reduced Hilbert space spanned by $ \ket{\phi_{\text{u,l}}}_{i}$. 

As in panel (b) in Fig.~\ref{fig5.fig}, 
we can find that the level crossing with respect to the ground state emerges twice while annealing.  
It might be expected as the emergence of the double first order phase transition while annealing discussed in Ref.~\cite{Seki2019}, 
where the system may come back to the ground state at the end of the annealing. 
However, there is no Landau--Zener tunneling between them, because we find no matrix elements between these level crossing states. 
In the case where $\Im \omega \neq 0$, non-zero value of matrix elements emerges between the Hilbert space projected by $\hat P$ and their orthogonal complement, which provides the anti-crossing of energy levels. In this case, discussion of the quantum annealing will become more complicated than that in the present study. 
At non-zero temperature case, the Schr\"odinger dynamics employed in this paper is not applicable, 
and it will be important to consider lower energy states in the Hilbert space projected by $\hat P$ as well as those in their orthogonal complement.

\bibliographystyle{apsrev4-1}
\bibliography{QuantumAnnealing_WajnflaszPickModel.bib}

\end{document}